\documentclass[pra,letterpaper,aps,10pt,superscriptaddress,twocolumn,floatfix]{revtex4-1}

\usepackage{amsmath, upgreek, amssymb, graphicx, subfigure, paralist}
\usepackage[colorlinks, linkcolor=blue, citecolor=blue, urlcolor=blue, breaklinks=true]{hyperref}

\newcommand{\eref}[1]{Eq.~(\ref{#1})}

\newcommand{\fref}[1]{Fig.~\ref{#1}}

\newcommand{\ie}{i.e.}

\newcommand{\eg}{e.g.}

\newcommand{\rmd}{\text{d}}

\newcommand{\twovec}[2]{\begin{pmatrix}#1\\#2\end{pmatrix}}
\newcommand{\twotwomat}[4]{\begin{bmatrix}#1&#2\\#3&#4\end{bmatrix}}

\newcommand{\abs}[1]{\lvert#1\rvert}

\newcommand{\im}[1]{\,\text{Im}\!\left\{#1\right\}}
\newcommand{\re}[1]{\,\text{Re}\!\left\{#1\right\}}

\makeatletter
\newlength \figurewidth
\makeatother

\begin{document}

\title{Strong coupling and long-range collective interactions in optomechanical arrays}

\date{\today}

\author{Andr\'e Xuereb}
\email[Corresponding author. ]{andre.xuereb@qub.ac.uk}
\affiliation{Centre for Theoretical Atomic, Molecular and Optical Physics, School of Mathematics and Physics, Queen's University Belfast, Belfast BT7\,1NN, United Kingdom}
\author{Claudiu Genes}
\affiliation{Institut f\"ur Theoretische Physik, Universit\"at Innsbruck, Technikerstrasse 25, A-6020 Innsbruck, Austria}
\affiliation{University of Vienna, Vienna Center for Quantum Science and Technology (VCQ), Faculty of Physics, Boltzmanngasse 5, 1090 Vienna, Austria}
\affiliation{ISIS (UMR 7006), Universit\'e de Strasbourg, Strasbourg, France}
\author{Aur\'elien Dantan}
\affiliation{QUANTOP, Danish National Research Foundation Center for Quantum Optics, Department of Physics and Astronomy, University of Aarhus, 8000 Aarhus C, Denmark}

\begin{abstract}
We investigate the collective optomechanics of an ensemble of scatterers inside a Fabry--P\'erot resonator and identify an optimized configuration where the ensemble is transmissive, in contrast with the usual reflective optomechanics approach. In this configuration, the optomechanical coupling of a specific collective mechanical mode can be several orders of magnitude larger than the single-element case, and long-range interactions can be generated between the different elements since light permeates throughout the array. This new regime should realistically allow for achieving strong single-photon optomechanical coupling with massive resonators, realizing hybrid quantum interfaces, and exploiting collective long-range interactions in arrays of atoms or mechanical oscillators.
\end{abstract}

\maketitle

The field of optomechanics has made tremendous progress over the past decades~\cite{Kippenberg2008,*Aspelmeyer2010}, cooling of massive mechanical oscillators to the motional quantum ground state being but one of a series of achievements that demonstrate the power of coupling light to moving scatterers~\cite{Chan2011,Teufel2011,*Verhagen2012}. The control of mechanical motion in the quantum regime has many important applications, ranging from precision measurements~\cite{Rugar2004}, quantum information processing~\cite{Mancini2003,*Pinard2005,*Stannigel2010}, and fundamental tests of quantum mechanics~\cite{Marshall2003,*Vitali2007,*RomeroIsart2011b,*RomeroIsart2011c,*Pikovski2012}, to the photonics sciences~\cite{Li2008,*Li2009b}. Despite recent progress the coupling between a single photon and a single phonon remains typically very weak, therefore necessitating the use of many photons to amplify the interaction~\cite{Kippenberg2008,*Aspelmeyer2010,Groblacher2009a}. In this regime, which is useful for cooling and light--motion entanglement generation, a stronger coupling per photon is desirable to limit the negative effects of using large powers, e.g., bulk temperature increases or phase-noise heating~\cite{Rabl2009}. Ultimately, reaching the strong (single-photon) coupling regime, in which a single quantum of light can appreciably affect the motion of the mechanical oscillator, is essential to exploiting fully the quantum nature of the optomechanical interaction, as exhibited by such effects as the optomechanical photon blockade~\cite{Rabl2011} and non-Gaussian mechanical states~\cite{Ludwig2008,*Nunnenkamp2011}.
\par
Among the various approaches currently followed to couple mechanical oscillators with optical resonators, a successful one involves positioning reflecting objects -- dielectric membranes~\cite{Thompson2008,*Wilson2009,Sankey2010,*Biancofiore2011,*Karuza2011}, atoms~\cite{Kruse2003,*Murch2008,*Brennecke2008,*Schleier-Smith2011}, or microspheres~\cite{Chang2009b,*RomeroIsart2010,*Monteiro2012} -- inside an optical cavity. With dielectric membranes the optomechanical interaction strength saturates to a fundamental limit $g$ as the reflectivity of the membrane approaches unity~\cite{Thompson2008,*Wilson2009}. For a highly reflective membrane placed to the center of a Fabry--P\'erot (FP) resonator of length $L$ and resonance frequency $\omega$, the single-photon coupling strength is given by the shift in cavity frequency when the mirror moves through a distance equal to the spread $x_0$ of its zero-point fluctuations, $g=2\omega x_0/L$, and is typically rather weak for a macroscopic cavity~\cite{Thompson2008,*Wilson2009}. Several approaches can be followed to improve quantum motional control in single membrane systems, by \eg, tailoring of the optical and mechanical properties of the individual membranes~\cite{Kemiktarak2012,*Bui2012,*Jockel2011}, using photothermal cooling forces~\cite{Usami2012}, active thermal noise compensation~\cite{Zhao2012}, optical trapping~\cite{Ni2012} techniques, or coupling to cold atoms~\cite{Hammerer2009,*Camerer2011,*Genes2011}.
\par
Another promising approach consists in exploiting \emph{collective} optomechanical interactions using microscopic ensembles of cold atoms~\cite{Kruse2003,*Murch2008,*Brennecke2008,*Schleier-Smith2011} or arrays of macroscopic mechanical oscillators~\cite{Hartmann2008,Bhattacharya2008a,Chang2011,Heinrich2011}. In the former the optomechanical coupling strength usually scales as $N^{1/2}$ with the number, $N$, of atoms, and the weakly-coupled atomic systems are said to demonstrate infinitely long-ranged interactions~\cite{Kruse2003,*Murch2008,*Brennecke2008,*Schleier-Smith2011}. In the latter, one can confine the light in periodic structures at the wavelength scale. In this vein, \eg, optomechanical crystals~\cite{Eichenfield2009} have proven to be very successful at obtaining large coupling strengths by decreasing the length of the effective cavity~\cite{Eichenfield2009b,Chan2011}. However, interactions between distant elements in arrays of massive scatterers are believed to be strongly suppressed~\cite{Heinrich2011}.\\
We provide here a unifying formalism that shows these two systems as limiting cases of a more generic model for an array of scatterers in a FP cavity. This allows us to identify an optimized configuration for the scatterers that is transmissive instead of the typical reflective approach. We base our treatment on the observation that around a transmission point of the mechanical system, the cavity response to a certain collective mechanical oscillation can be greatly enhanced. This is to be seen as an alternative to the traditional approach that requires smaller cavities (on the order of a few wavelengths~\cite{Eichenfield2009}) to increase the field density of modes and thus the coupling between light and mechanics. Our analysis reveals a regime where, regardless of whether the scatterers are atoms or mobile dielectrics, the coupling strength (i)~scales superlinearly ($\varpropto N^{3/2}$) with the number of scatterers and (ii)~does not saturate as the reflectivity of the elements approaches unity. This allows in principle multi-element opto- or electro-mechanical systems to reach single-photon optomechanical coupling strengths orders of magnitude larger than those currently possible, and does not require wavelength-scale confinement of the light field. Concomitantly, we show that in this configuration (iii)~the resonator field couples to a specific collective mechanical mode supporting inter-element interactions that are as long-ranged as the array itself. Since the model is applicable both to mobile dielectrics, such as membranes~\cite{Thompson2008,*Wilson2009} or microspheres~\cite{Chang2009b,*RomeroIsart2010,*Monteiro2012}, and to cold atoms in an optical lattice~\cite{Birkl1995,*Weidemuller1995,Schilke2011}, this new regime should realistically allow for achieving strong single-photon optomechanical coupling and realizing quantum optomechanical interfaces. It also opens up avenues for the exploitation and engineering of long-ranged cooperative interactions in optomechanical arrays.

\begin{figure}[t]
 \includegraphics[width=\figurewidth]{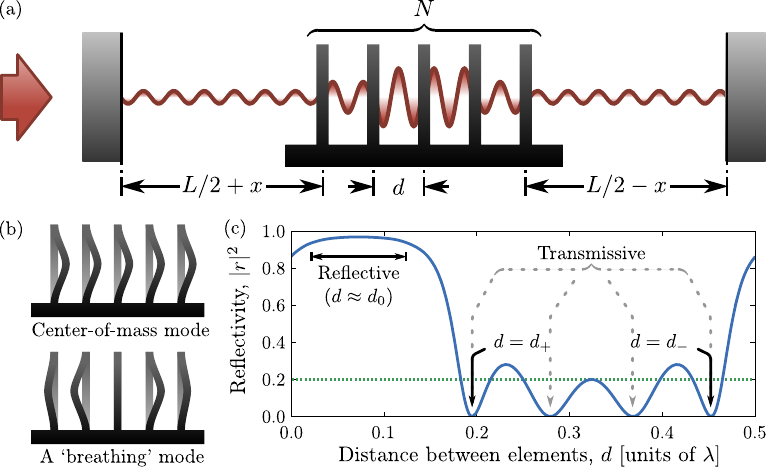}
\caption{Schematic of model, motional modes, and working points. (a)~System considered:\ $N$ equidistant elements positioned in the field of a Fabry--P\'erot resonator (we shall consider only the case for which $L\gg Nd$). (b)~Two examples of collective motional modes:\ the center-of-mass mode, and an example of a `breathing' modes. (c)~Free-space reflectivity of a $5$-element array as a function of the element spacing $d$; the curve is periodic with period $\lambda/2$. The intensity reflectivity of each element is 20\% (dotted green line).}
 \label{fig:Scheme}
\end{figure}

Let us again consider a lossless membrane, of thickness smaller than a wavelength, placed inside a FP resonator. This time, we suppose that the membrane has an amplitude reflectivity $r$, which we parametrize in terms of the polarizability $\zeta\equiv-\abs{r}\big/\sqrt{1-\abs{r}^2}$. The single-photon optomechanical coupling strength is now $g_0=g\abs{r}$, which is maximized to $g$ for large $\lvert\zeta\rvert$, \ie, in the \textit{reflective} regime $\abs{r}\to1$. In order to illustrate the emergence of collective optomechanics, we consider now two identical membranes placed symmetrically in the resonator at a distance $d$ from each other, in the spirit of \fref{fig:Scheme}(a) and Ref.~\cite{Bhattacharya2008a}. As we justify below, the effective polarizability of the two-element system is found to be of the form $\chi=2\zeta[\cos(kd)-\zeta\sin(kd)]$ for light having wavenumber $k$ (wavelength $\lambda$). This effective polarizability, and thereby the reflectivity, vanishes when $d$ is chosen such that $kd=\tan^{-1}(1/\zeta)\mod\pi$. Assuming this \textit{transmissive} condition, one can linearize the cavity resonance condition for a small variation $\delta d$ of the mirror spacing. This readily gives an optomechanical coupling strength
\begin{equation}
\label{eq:TwoMembraneg}
g_0^\prime=\biggl|\frac{\delta\omega}{\delta d}\biggr|x_0^\prime\approx\sqrt{2}\,g\frac{\abs{r}}{1-\abs{r}}\,,
\end{equation}
provided $d|\zeta|^2\ll L$, where $x_0^\prime=x_0/\sqrt{2}$ is the extent of the zero-point motion for this breathing mode. It is evident that $g_0^\prime$ scales more favorably with $\abs{r}$ than $g_0$. One can interpret this result by noting that as the reflectivity of the individual elements is increased, the constructive interference that is responsible for making the array transmissive also strongly enhances the dispersive response of the cavity around this working point. In a symmetric situation the displacement of a mirror in one direction will cause the field to adjust so that the other mirror moves in the opposite direction, thereby balancing the power impinging on the two mirrors. In this simple two-element case, the radiation-pressure force thus couples naturally to a breathing mode [\fref{fig:Scheme}(b)].

\begin{figure}[t]
 \centering
 \includegraphics[width=\figurewidth]{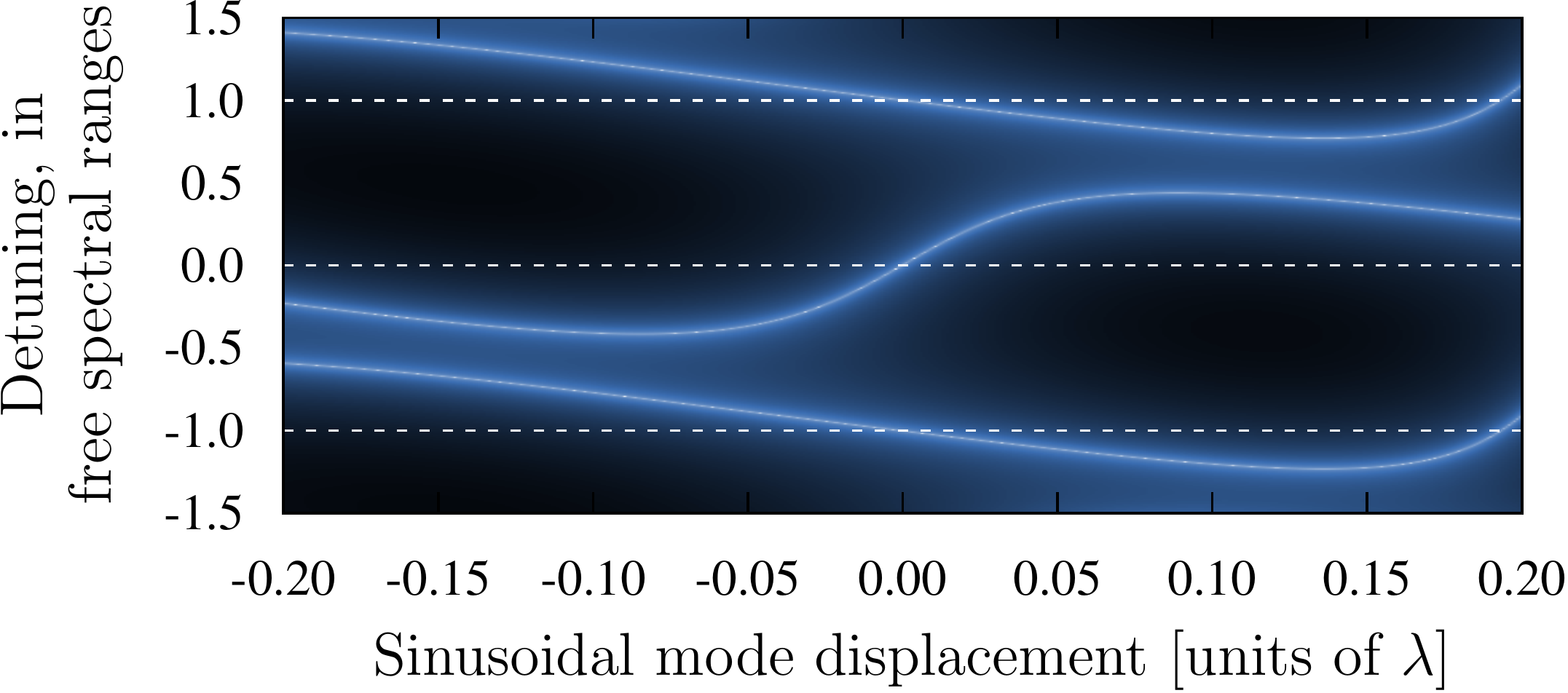}
\caption{Transmission through a cavity with 5 immobile elements. The dashed lines denote the unperturbed resonances, which are shifted when the displacement of the sinusoidal mode, a type of breathing mode [\fref{fig:Scheme}(b)] that is defined in the text, is nonzero. ($\zeta=-0.5$, $L\approx6.3\times10^4\lambda$, $d=d_-$, bare cavity finesse\ $\approx\,3\times10^4$.)}
 \label{fig:Transmission}
\end{figure}

\section{Optical `superscatterers'}
To treat the general case of an array of $N$ equally-spaced elements in free space we make use of the transfer matrix formalism~\cite{Deutsch1995} for one-dimensional systems of polarizable scatterers, and derive the response of the system to a propagating light field. As is well-known from the theory of dielectric mirrors, the reflectivity of the ensemble can be tuned to have markedly different behaviors at a given frequency [\fref{fig:Scheme}(c)]. An array of $N$ equally-spaced identical elements, each of polarizability $\zeta$, can be described through a matrix that relates left- and right-propagating fields on either side of the array~(see Appendix). For real $\zeta$, \emph{$N$ lossless scatterers behave as a collective `superscatterer'} having effective polarizability $\chi=\zeta\,\sin[N\cos^{-1}(a)]\big/\sqrt{1-a^2}$, with $a=\cos(kd)-\zeta\sin(kd)$, together with a phase shift $\mu$, which is the phase accrued on reflection from the stack. The ensemble attains its largest reflectivity for $kd=kd_0\equiv-\tan^{-1}(\zeta)$, $\chi=\chi_0\equiv-i\sin\bigl[N\cos^{-1}\bigl(\sqrt{1+\zeta^2}\big)\bigr]$, and becomes fully transmissive ($\chi=0$) for $kd=kd_\pm\equiv-\tan^{-1}(\zeta)\pm\cos^{-1}\bigl[(1+\zeta^2)^{-1/2}\cos(\pi/N)\bigr]$. For absorbing scatterers, setting $d=d_-$ (modulo $\lambda/2$) helps minimize the effects of absorption~(see Appendix). These working points are illustrated in \fref{fig:Scheme}(c). From this point onwards, the array can be treated as a single scatterer, keeping in mind the dependence of $\chi$ on the inter-element spacing.

\begin{figure}[t]
 \centering
 \includegraphics[width=\figurewidth]{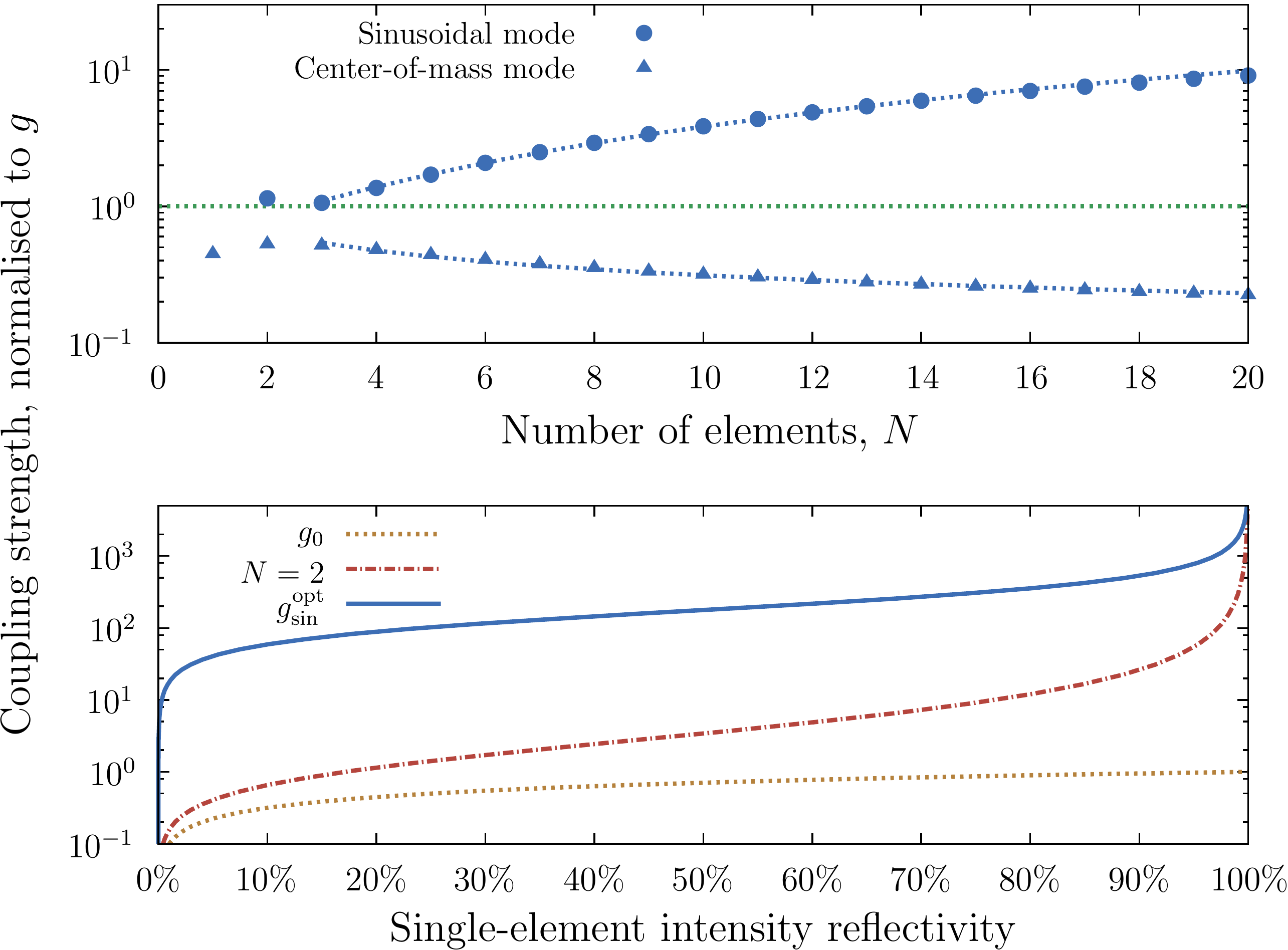}
\caption{Coupling strengths for multi-element arrays. These curves are scaled to $g\approx2\pi\times15$\,Hz (dotted green line), which is the upper bound for the reflective case. \emph{Top:}~Scaling with $N$ of the normalized coupling strength for the sinusoidal ($\varpropto N^{3/2}$) and center-of-mass ($\varpropto N^{-1/2}$) modes, as illustrated by the dotted curves. ($\zeta=-0.5$, $L\approx6.3\times10^4\lambda$, $d=d_-+20\,\lambda$.) \emph{Bottom:}~Optimized sinusoidal coupling $g_\mathrm{sin}^\mathrm{opt}$ compared to the coupling for $N=2$, demonstrating the collectively-enhanced coupling strength, and to $g_0$. ($d=d_-$.)}
 \label{fig:Coupling}
\end{figure}

\section{Ensemble coupling strength}
\begin{figure}[t]
 \centering
 \includegraphics[width=\figurewidth]{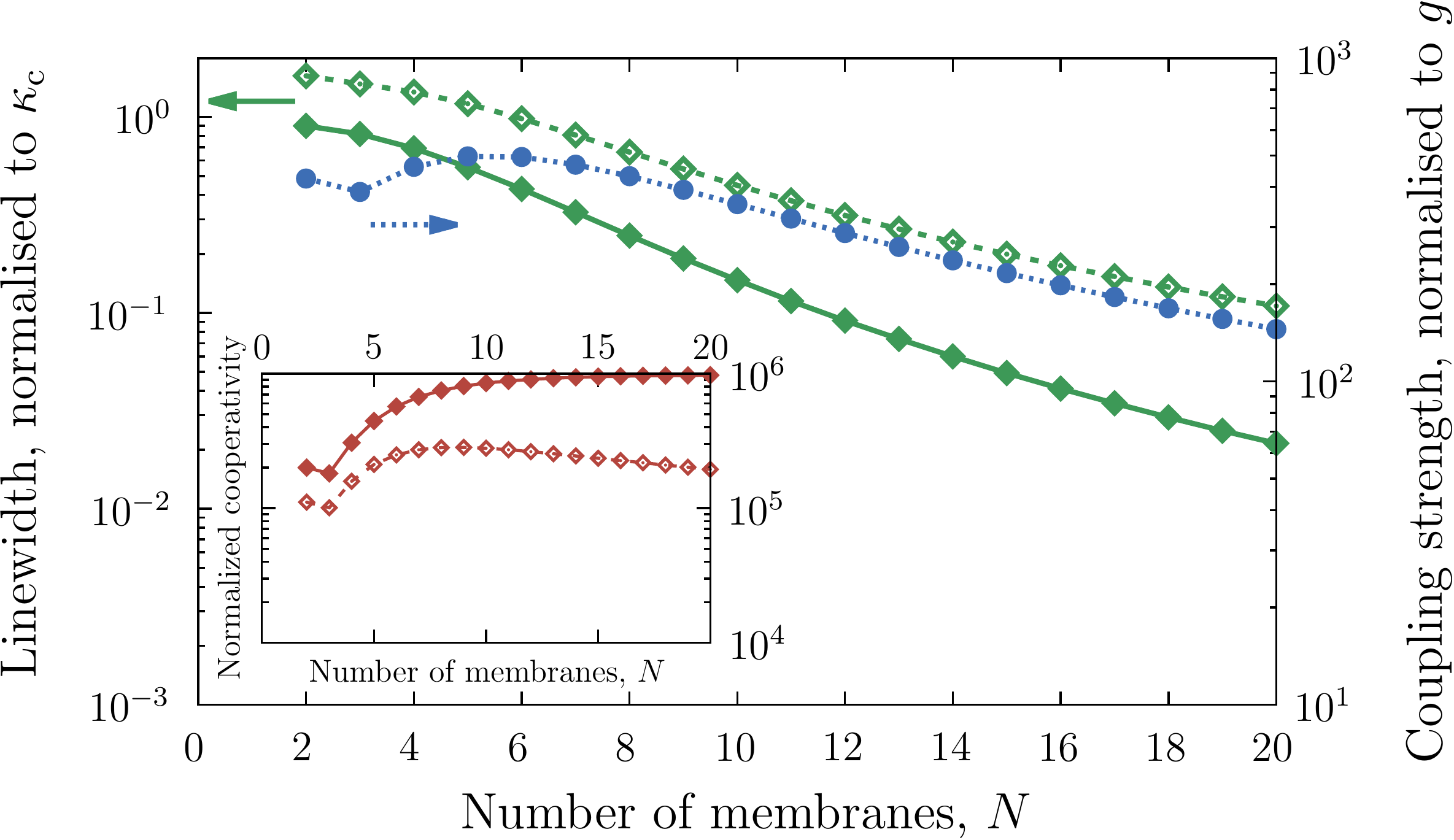}
\caption{Effective cavity linewidth $\kappa_\mathrm{eff}$ (diamonds) and optomechanical coupling strength $g_\mathrm{sin}$ (cf.\ \fref{fig:Coupling}; circles) as a function of the number of membranes. \emph{Inset:}~Cooperativity $g_\mathrm{sin}^2/(\kappa_\mathrm{eff}\Gamma_\mathrm{dec})$ normalized to the single-element cooperativity $g^2/(\kappa_\mathrm{c}\Gamma_\mathrm{dec})$; the non-normalized cooperativity can reach values $>10$ (see text). Closed symbols represent absorption-free membranes, open symbols an absorption of $10^{-5}$ per membrane. ($99.4$\% reflectivity, $d=d_-+10\lambda$, $L\approx6.3\times10^4\lambda$, bare cavity finesse $\approx\,3\times10^4$.)}
 \label{fig:Cooperativity}
\end{figure}

When placed inside a cavity, at first neglecting any motion, this array of scatterers modifies the resonance condition, such that the resonances of the system are given by the solutions to~(see Appendix)
\begin{equation}
\label{eq:ResonanceCondition}
e^{ikL}=\frac{e^{-i\mu}}{1+i\chi}\biggl[i\chi\cos(2kx)\pm\sqrt{1+\chi^2\sin^2(2kx)}\biggr]\,,
\end{equation}
where $x$ is the displacement of the ensemble with respect to the cavity center, and $\mu\equiv\mu(x_1,x_2,\dots)$ and $\chi\equiv\chi(x_1,x_2,\dots)$ depend on the positions $x_j$ of the individual elements. For a particular configuration, \eref{eq:ResonanceCondition} is solved numerically to find the resonance frequency $\omega=kc$. A small shift $\delta x_j$ in the position of the $j$\textsuperscript{th} element in the array shifts this resonance:\ $\omega\to\omega-g_j\delta x_j$. The vector $(g_j)$ defines the profile of the collective motional mode that is coupled to the cavity field. In the case of a transmissive ensemble the intensity profile peaks at the center of the array [\fref{fig:Scheme}(a)]. The optomechanical coupling strength $g_j$ for the $j$\textsuperscript{th} membrane is strongest where the difference in amplitudes across the membrane is greatest, $j\approx(N+2)/4$ or $(3N+2)/4$, resulting in $g_j\varpropto\sin[\pi(2j-1)/N]$ and a mechanical mode whose profile varies sinusoidally along the array. In \fref{fig:Transmission} we plot the transmission of the cavity ($\mathcal{T}_\mathrm{cav}$~(see Appendix)) as a function of frequency and the displacement of this sinusoidal mode. The dashed lines represent solutions to \eref{eq:ResonanceCondition}, \ie, in the absence of membrane motion, and are one free-spectral range apart. The gradient of the bright curves at any point is a direct measure of the optomechanical coupling strength for the sinusoidal mode at that point. The center of the plot corresponds to our working point; the adjacent optical resonances are to a good approximation one bare-cavity free-spectral range apart. In the situations we consider here, we have checked that the linear coupling largely dominates over the quadratic coupling~(see Appendix).\\
Generically, one obtains the linear optomechanical coupling strength by linearizing \eref{eq:ResonanceCondition} about one of its solutions. For a center-of-mass motion [cf.\ \fref{fig:Scheme}(b)] in the reflective regime, $d=d_0$, we thus obtain $g_\mathrm{com}=g\sqrt{\mathcal{R}/N}$, where $\mathcal{R}=\chi_0^2\big/\bigl(1+\chi_0^2\bigr)$ is the maximal intensity reflectivity of the ensemble. As $N$ or $\zeta$ increase, $\mathcal{R}$ saturates to $1$ and $g_\mathrm{com}$ scales as $N^{-1/2}$. This scaling can be explained simply by noting that the motional mass $m_N$ of $N$ elements is $N$ times that of a single one; the single-photon coupling strength, which is proportional to $1/\sqrt{m_N}$, therefore \emph{decreases} with $N$.\\
In the transmissive regime, $d=d_-$ (modulo $\lambda/2$), then, the cavity field couples to the sinusoidal mode with a collective coupling strength (for large $N$~(see Appendix))
\begin{equation}
\label{eq:SinusoidalCouplingSimplified}
g_\mathrm{sin}=\frac{\tfrac{\sqrt{2}}{\pi}\,g\,\zeta^2N^{3/2}}{1+\tfrac{2}{\pi^2}\tfrac{d}{L}\zeta^2N^3}\approx\frac{\sqrt{2}}{\pi}\,g\,\zeta^2N^{3/2}\,,
\end{equation}
the last expression being valid for $L/d\gg2\zeta^2N^3/\pi^2$. Optimizing over $N$ for arbitrary $L/d$, we obtain $g_\mathrm{sin}^\mathrm{opt}=\tfrac{1}{2}g\sqrt{L/d}\abs{\zeta}$~\footnote{This expression is valid for $\abs{\zeta}$ that is not too large, since the optimal number of elements must be $\geq2$.}. This favorable scaling with both $N$ and $\abs{r}$, as shown in \fref{fig:Coupling}, is a significant improvement over the state of the art. Close inspection reveals that $g_\mathrm{sin}^\mathrm{opt}$ is proportional to $1\big/\sqrt{Ld}$ and therefore can be improved either by making the main cavity smaller (\ie, decreasing $L$) or, independently, by positioning the elements closer together (decreasing $d$). The effect we describe is therefore qualitatively different from constructing a smaller cavity having dimensions on the order of $\lambda$~\cite{Eichenfield2009b}, and also provides a practical route towards integrating strongly-coupled optomechanical systems with, \eg, ensembles of atoms in the same cavity~\cite{Genes2011}.\\
An interesting effect arises in the regime where $g_\mathrm{sin}$ saturates and eventually starts decreasing as a function of $N$; the scatterers then act to narrow the cavity resonance substantially. This arises from an effective lengthening of the cavity, due to the presence of the array, to a length $L_\mathrm{eff}\equiv L+\tfrac{2}{\pi^2}d\zeta^2N^3$. Since the cavity finesse in the transmissive regime is fixed by the end mirrors, it follows that the linewidth of the cavity is $\kappa_\mathrm{eff}\varpropto1/L_\mathrm{eff}$ (bare cavity linewidth $\kappa_\mathrm{c}\varpropto1/L$), which has possible applications in hybrid systems along the same lines as those of electromagnetically-induced transparency in Ref.~\cite{Genes2011}. When using low-finesse cavities and low mechanical oscillation frequencies, this effect could be used to place the system well within the sideband-resolved regime. As shown in \fref{fig:Cooperativity}, $g_\mathrm{sin}$ and $\kappa_\mathrm{eff}$ compete to give rise to a constant cooperativity $g_\mathrm{sin}^2/(\kappa_\mathrm{eff}\Gamma_\mathrm{dec})$ for large $N$ ($1/\Gamma_\mathrm{dec}$ is the mechanical decoherence timescale, assumed independent of $N$~(see Appendix)). In the presence of absorption, which ultimately limits the linewidth narrowing, there exists an optimum number of elements which maximizes the cooperativity to a value that can still be several orders of magnitude larger than the single-element cooperativity.

\emph{Long-range collective interactions.}---The collective nature of the interaction that is responsible for these large coupling strengths also gives rise to an effective `non-local' interaction between the scatterers, where the motion of any particular element influences greatly elements further away, and not just its nearest neighbors. In the simplest picture of a weak linearized optomechanical interaction~\cite{Kippenberg2008,*Aspelmeyer2010} in which the field is adiabatically eliminated, the interaction Hamiltonian is proportional to $\sum_{l,j} g_lg_j\hat{x}_l\hat{x}_j$, and mediates a \emph{macroscopically long-ranged} effective interaction between pairs of elements (position operators $\hat{x}_l$ and $\hat{x}_j$). By contrast, in the reflective regime the light does not permeate through the ensemble, and the inter-element mechanical interactions would therefore be correspondingly short-ranged (see, for example, Ref.~\cite{Heinrich2011}). Transmissive arrays with well-designed spacings and polarizabilities could be used to engineer specific optomechanical interactions and gain insight into collective optomechanics phenomena~\cite{Chang2011,Heinrich2011}.

\emph{Numerical example, tolerance to imperfections.}---The power of this approach to optomechanics is best seen through a numerical illustration. If we take commercial silicon nitride membranes~\cite{Thompson2008} with an intensity reflectivity of 20\% ($\zeta=-0.5$) and $x_0=1.8$\,fm, and a cavity with $L=6.7$\,cm and a wavelength of $1064$\,nm, we can estimate $g_\mathrm{com}\approx2\pi\times(12.8\times N^{-1/2}$\,Hz$)$ for $N\gtrsim3$. For the sinusoidal mode, and with the same parameters, $g_\mathrm{sin}\approx2\pi\times(1.3\times N^{3/2}$\,Hz$)$ for large $N$; an improvement by over an order of magnitude when $N=10$ (cf.\ \fref{fig:Coupling}). A transparent ensemble potentially provides a much stronger optomechanical coupling than a reflective one; indeed $g_\mathrm{sin}/g_\mathrm{com}\varpropto N^2$. Let us now consider highly-reflective membranes~\cite{Kemiktarak2012} having $99.4$\% intensity reflectivity ($\zeta=-12.9$), $x_0=2.7$\,fm, and $\omega_\mathrm{m}=2\pi\times211$\,kHz. For a $0.25$\,cm-long cavity with finesse $F=1.2\times10^5$, $d=d_-$, and $N=5$ membranes, one obtains $g_\mathrm{com}\approx2\pi\times600$\,Hz and $g_\mathrm{sin}\approx2\pi\times270$\,kHz, which is larger than both $\omega_\mathrm{m}$ and $\kappa_\mathrm{c}=2\pi\times250\,$kHz. At a temperature of $1$\,K and with a mechanical quality factor of $10^6$ the \emph{single photon} cooperativity is ca.\ $14$ for this system; strong coupling between a single photon and a single phonon is already within reach with only a few elements. A thorough numerical investigation~(see Appendix) reveals that our results are robust with respect to various experimentally-relevant deviations from the idealized system considered here, such as the errors in the positioning of the individual membranes, and non-uniform membrane reflectivity or absorption. For example, for $N=5$ and $\zeta=-0.5$, the numerically-calculated coupling strength typically lies within $12$\% of the above value for position fluctuations of $\pm10$\,nm, inhomogeneities in $\zeta$ of $\pm10$\%, and absorption per element of $\gtrsim10^{-3}$.\\
Moving away from highly-reflective scatterers, we can apply our results to systems of very low reflectivity, such as atoms, molecules, dielectric microspheres, etc. It should be first noted that all these systems have a reflectivity on the order of $10^{-6}$, which means that $N\lvert\zeta\rvert\ll1$ in typical experiments (for example with cold atoms in cavities~\cite{Kruse2003,*Murch2008,*Brennecke2008,*Schleier-Smith2011}), \ie, the particles do not significantly modify the mode structure of the cavity resonance. In this case $g_\mathrm{sin}$ reduces to the expected $N^{1/2}$ scaling that arises from the independent coupling of well-localized scatterers interacting with an unmodified cavity field~\cite{Kruse2003,*Murch2008,*Brennecke2008,*Schleier-Smith2011}. We note, however, that recent experiments~\cite{Schilke2011} using ``pancake''-shaped clouds of cold atoms in an optical lattice have shown intensity reflectivities as high as $80$\% and are approaching a regime where the effects discussed previously may be observed.

\section{Conclusions}
We have made use of a fully analytical theory to explore novel interactions between the collective mechanical dynamics of an array of equidistant scatterers inside a cavity, and the cavity field itself. Our ideas apply generically across a wide range of systems; any system that can be modeled as a one-dimensional chain of scatterers (\eg, membranes, atoms~\cite{Schilke2011}, optomechanical crystals~\cite{Chan2011}, or dielectric microspheres~\cite{Chang2009b,*RomeroIsart2010,*Monteiro2012}) is amenable to a similar analysis and shows the same rich physics. Similar methods would allow the extension of these ideas to more complicated systems where the polarizability is a function of frequency or of position along the array, or systems involving the interaction of arrays of refractive elements with multiple optical modes.

\section{Acknowledgements}
We acknowledge support from the Royal Commission for the Exhibition of 1851 (A.X.), the NanoSci-E+ Project ``NOIs'' and the Austrian Science Fund (FWF) (C.G.), and the EU CCQED and PICC projects (A.D.). We would also like to thank J.\ Bateman, K.\ Hammerer, I.\ D.\ Leroux, M.\ Paternostro, and H.\ Ritsch for fruitful discussions.

\appendix
\section{Transfer matrix for $N$-membrane stack}
Let us consider $N$ equally-spaced identical non-absorbing membranes, each of which has a thickness much smaller than an optical wavelength. The spacing $d$ between these membranes determines the overall optical properties of the ensemble. We shall treat the system as a strictly one-dimensional system, and we shall use the transfer matrix formalism~\cite{Deutsch1995,Xuereb2009b}. Our starting point is the matrix that links the fields interacting with a single membrane of polarisability $\zeta$ ($\zeta\in\mathbb{R}$ for a lossless membrane),
\begin{equation}
M_\mathrm{m}(\zeta)\equiv\twotwomat{1+i\zeta}{i\zeta}{-i\zeta}{1-i\zeta}\,,
\end{equation}
and the matrix that describes propagation of a monochromatic beam of wavenumber $k$ over a distance $d$ through free space,
\begin{equation}
M_\mathrm{p}(d)\equiv\twotwomat{e^{ikd}}{0}{0}{e^{-ikd}}\,,
\end{equation}
both of which have unit determinant. We wish to evaluate a product of the form
\begin{equation}
M_\mathrm{m}(\zeta)\cdot M_\mathrm{p}(d)\cdot M_\mathrm{m}(\zeta)\cdots M_\mathrm{m}(\zeta)\,,
\end{equation}
with $N$ copies of $M_\mathrm{m}(\zeta)$. First, we note that
\begin{multline}
M_\mathrm{p}(d/2)\cdot M_\mathrm{m}(\zeta)\cdot M_\mathrm{p}(d)\cdots M_\mathrm{m}(\zeta)\cdot M_\mathrm{p}(d/2)\\
=\bigl[M_\mathrm{p}(d/2)\cdot M_\mathrm{m}(\zeta)\cdot M_\mathrm{p}(d/2)\bigr]^N\equiv M^N\,,
\end{multline}
where the second line defines the matrix $M$:
\begin{equation}
M\equiv\twotwomat{(1+i\zeta)e^{ikd}}{i\zeta}{-i\zeta}{(1-i\zeta)e^{-ikd}}\,.
\end{equation}
We can easily see that $\det M=1$, whereby it can be shown~\cite{Yeh2005} that for real $\zeta$ we can write
\begin{equation}
M^N=\twotwomat{(1+i\chi)e^{i(kd+\mu)}}{i\chi}{-i\chi}{(1-i\chi)e^{-i(kd+\mu)}}\,,
\end{equation}
where $\chi\equiv\zeta U_{N-1}(a)$, with $U_n(x)$ being the $n$\textsuperscript{th} Chebyshev polynomial of the second kind, $a=\cos(kd)-\zeta\sin(kd)$, and
\begin{equation}
e^{i\mu}=\frac{1-i\zeta
U_{N-1}(a)}{(1-i\zeta)U_{N-1}(a)-e^{ikd}U_{N-2}(a)}\,.
\end{equation}
Upon removing the padding of $d/2$ from either side, we obtain the matrix that describes the $N$-membrane ensemble:
\begin{equation}
M_N\equiv M_\mathrm{p}[\mu/(2k)]\cdot M_\mathrm{m}(\chi)\cdot M_\mathrm{p}[\mu/(2k)]\,.
\end{equation}
Thus, as stated in the main text, $N$ lossless membranes behave as a collective `supermembrane' of polarisability $\chi$ along with a phase shift $\mu/2$ on either side of the stack.\par
Let us note, finally, the link between the transfer matrix of an optical system and its amplitude transmissivity. Indeed, suppose we can describe a system by means of the transfer matrix
\begin{equation}
\twotwomat{m_{11}}{m_{12}}{m_{21}}{m_{22}}\,.
\end{equation}
Then, the complex transmissivity of the system is simply
\begin{equation}
\label{eq:TransGeneral}
\mathcal{T}=\frac{1}{m_{22}}\,.
\end{equation}
This is used throughout this work to characterize the optical properties of our system.

\section{Compound system:\ membrane stack inside cavity}
Following the main text, let us now place our membrane stack inside a Fabry--P\'erot cavity of length $L$. The transfer matrix describing this system is then
\begin{multline}
M_\mathrm{cav}\equiv M_\mathrm{m}(Z)\cdot M_\mathrm{p}(L/2+x)\\
\cdot M_N\cdot M_\mathrm{p}(L/2-x)\cdot M_\mathrm{m}(Z)\,.
\end{multline}
Here $x$ is the displacement of the left edge of the ensemble with respect to the center of the cavity and $Z$ is the polarizability of the cavity mirrors, assumed equal for both. The transmission of the system, following \eref{eq:TransGeneral}, is given by
\begin{equation}
\mathcal{T}_\mathrm{cav}=\frac{1}{\bigl(M_\mathrm{cav}\bigr)_{22}}\,;
\end{equation}
the maxima of $\mathcal{T}_\mathrm{cav}$ give the resonances of this system. In order to find these resonances analytically, we consider a simpler system where the cavity mirrors are perfect; we need only solve the relation
\begin{multline}
\twovec{\phantom{+}1}{-1}\varpropto\twotwomat{e^{i\theta}}{0}{0}{e^{-i\theta}}\times\twotwomat{1+i\chi}{i\chi}{-i\chi}{1-i\chi}\\
\times\twotwomat{e^{i\phi}}{0}{0}{e^{-i\phi}}\cdot\twovec{\phantom{+}1}{-1}\,,
\end{multline}
with $\theta\equiv k(L/2+x)+\mu/2$ and $\phi\equiv k(L/2-x)+\mu/2$. We thus obtain
\begin{equation}
\label{eq:ResonanceSolution}
e^{ikL}=\frac{e^{-i\mu}}{1+i\chi}\biggl[i\chi\cos(2kx)\pm\sqrt{1+\chi^2\sin^2(2kx)}\biggr]
\end{equation}
However, we immediately see that this equation is transcendental in $k$, and therefore cannot be solved analytically; this equation is easily solvable for $L$, and we therefore know the resonant length of our cavity. It is now a legitimate question to ask: `If $d$ (or $x$) shifts by a small amount, how much will the resonant frequency of \emph{this} cavity shift?' This question is, of course, easily answered by expanding \eref{eq:ResonanceSolution} in small increments about its solution. Assuming a dominantly linear effect, we replace $k\to k_0+\delta k$, $x\to x+\delta x$, $\chi\to\chi+\delta\chi$, and $\mu\to\mu+\delta\mu$ in \eref{eq:ResonanceSolution}. Around resonance, the result simplifies to
\begin{align}
\label{eq:GeneratorEquation}
L\delta k+\delta\mu=&\ \biggl[-1\pm\cos(2k_0x)\Big/\sqrt{1+\chi^2\sin^2(2k_0x)}\biggr]\nonumber\\
&\qquad\qquad\times\delta\chi/\bigl(1+\chi^2\bigr)\nonumber\\
&\mp\biggl[2\chi\sin(2k_0x)\Big/\sqrt{1+\chi^2\sin^2(2k_0x)}\biggr]\nonumber\\
&\qquad\qquad\times(x\delta k+k_0\delta x)\,.
\end{align}

\section{Center-of-mass coupling strength}
Let us start from \eref{eq:GeneratorEquation}. For the center-of-mass motion, $\partial\mu=\partial\chi=0$, and we assume that $\abs{L/x}$ is very large, such that we can write
\begin{equation}
L\delta k=\mp\biggl[2\chi\sin(2k_0x)\Big/\sqrt{1+\chi^2\sin^2(2k_0x)}\biggr]k_0\delta x\,.
\end{equation}
The right-hand-side of this equation is maximized when $\sin(2k_0x)=\mp1$, whereby
\begin{equation}
L\delta k=2k_0\bigl(-\chi\big/\sqrt{1+\chi^2}\bigr)\delta x\,.
\end{equation}
This is, in absolute value, a monotonically-increasing function of $\abs{\chi}$ and is therefore maximized when $\chi$ attains its largest value, $\zeta\,U_{N-1}\bigl(\sqrt{1+\zeta^2}\bigr)$, which leads to $g_\mathrm{com}$ as defined in the main text.

\begin{figure}[b]
 \includegraphics[scale=0.5]{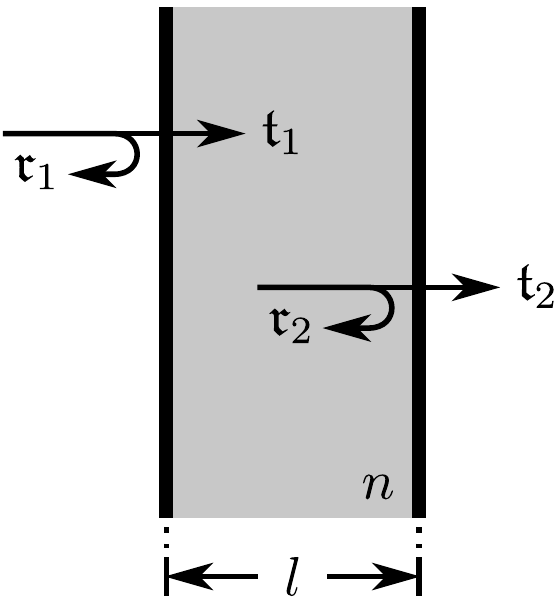}
\caption{Modeling an individual membrane as a plate of thickness $l$ and refractive index $n$.}
 \label{fig:Plate}
\end{figure}

\section{Coupling to each individual membrane}
The matrix $M_N$ representing the ensemble can be written, for $1\leq j\leq N$,
\begin{multline}
\twotwomat{e^{i\mu_1/2}}{0}{0}{e^{-i\mu_1/2}}\twotwomat{1+i\chi_1}{i\chi_1}{-i\chi_1}{1-i\chi_1}\\
\times\twotwomat{e^{i(\mu_1/2+\nu+k\delta x_j)}}{0}{0}{e^{-i(\mu_1/2+\nu+k\delta x_j)}}\\
\times\twotwomat{1+i\zeta}{i\zeta}{-i\zeta}{1-i\zeta}\twotwomat{e^{i(\mu_2/2+\nu-k\delta x_j)}}{0}{0}{e^{-i(\mu_2/2+\nu-k\delta x_j)}}\\
\times\twotwomat{1+i\chi_2}{i\chi_2}{-i\chi_2}{1-i\chi_2}\twotwomat{e^{i\mu_2/2}}{0}{0}{e^{-i\mu_2/2}}\,,
\end{multline}
where $\mu_1$ and $\chi_1$ describe the ensemble formed by the $n_1=j-1$ membranes to the `left' of the $j$\textsuperscript{th}, and $\mu_2$ and $\chi_2$ the one formed by the $n_2=N-j$ membranes to its `right'. The displacement of the $j$\textsuperscript{th} element is denoted $\delta x_j$; all other membranes are in their equilibrium position. In the transmissive regime, to lowest order in $k\delta x_j$ in each entry, the matrix product above can be written, with the above choice for $\nu$,
\begin{equation}
\twotwomat{e^{i\mu}+\alpha\,\delta x_j}{\beta\,\delta x_j}{\beta^\ast\,\delta x_j}{e^{-i\mu}+\alpha^\ast\,\delta x_j}\,,
\end{equation}
\begin{figure*}[t]
 \includegraphics[width=0.5\figurewidth]{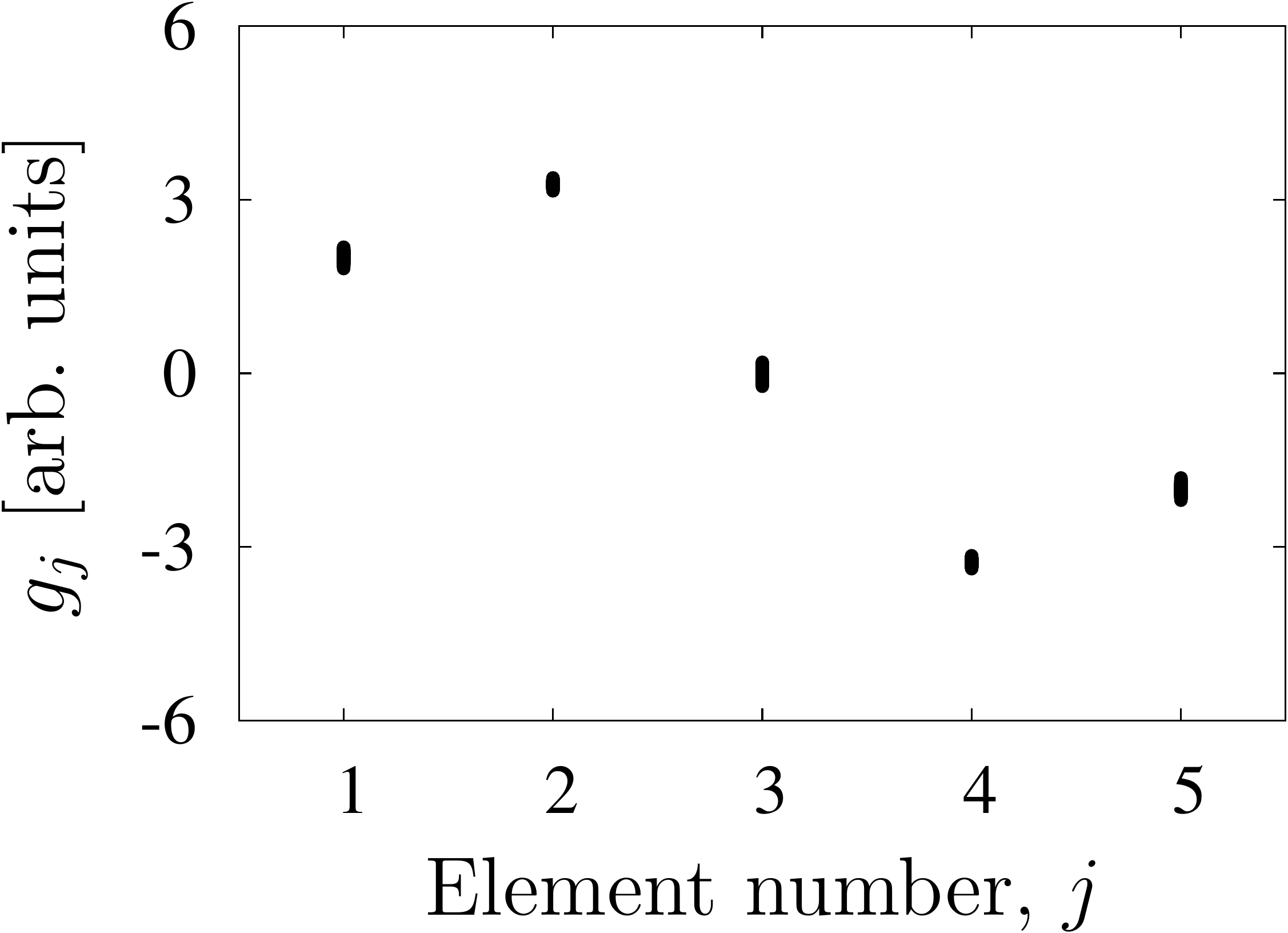}\quad\ 
 \includegraphics[width=0.5\figurewidth]{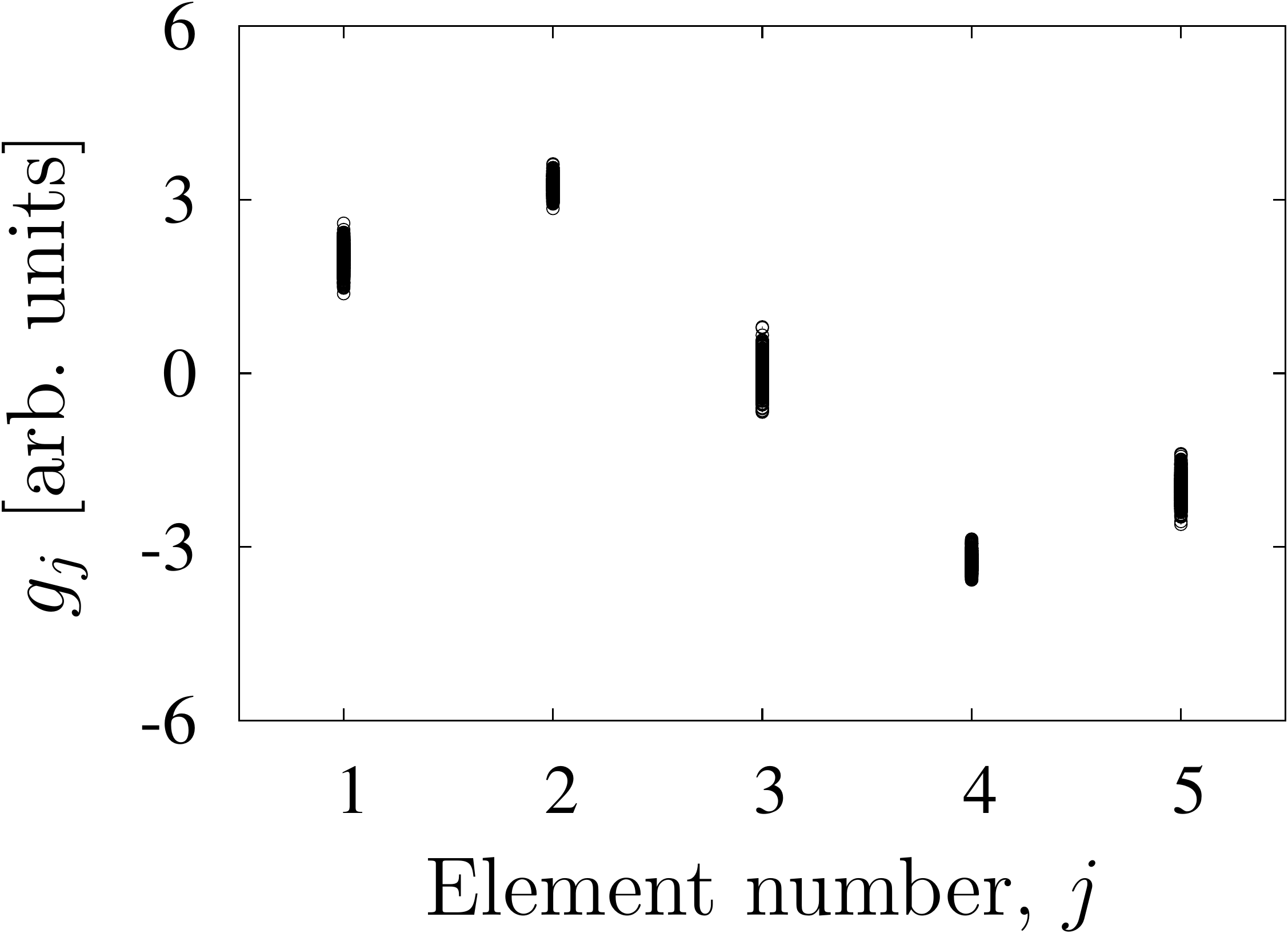}\quad\ 
 \includegraphics[width=0.5\figurewidth]{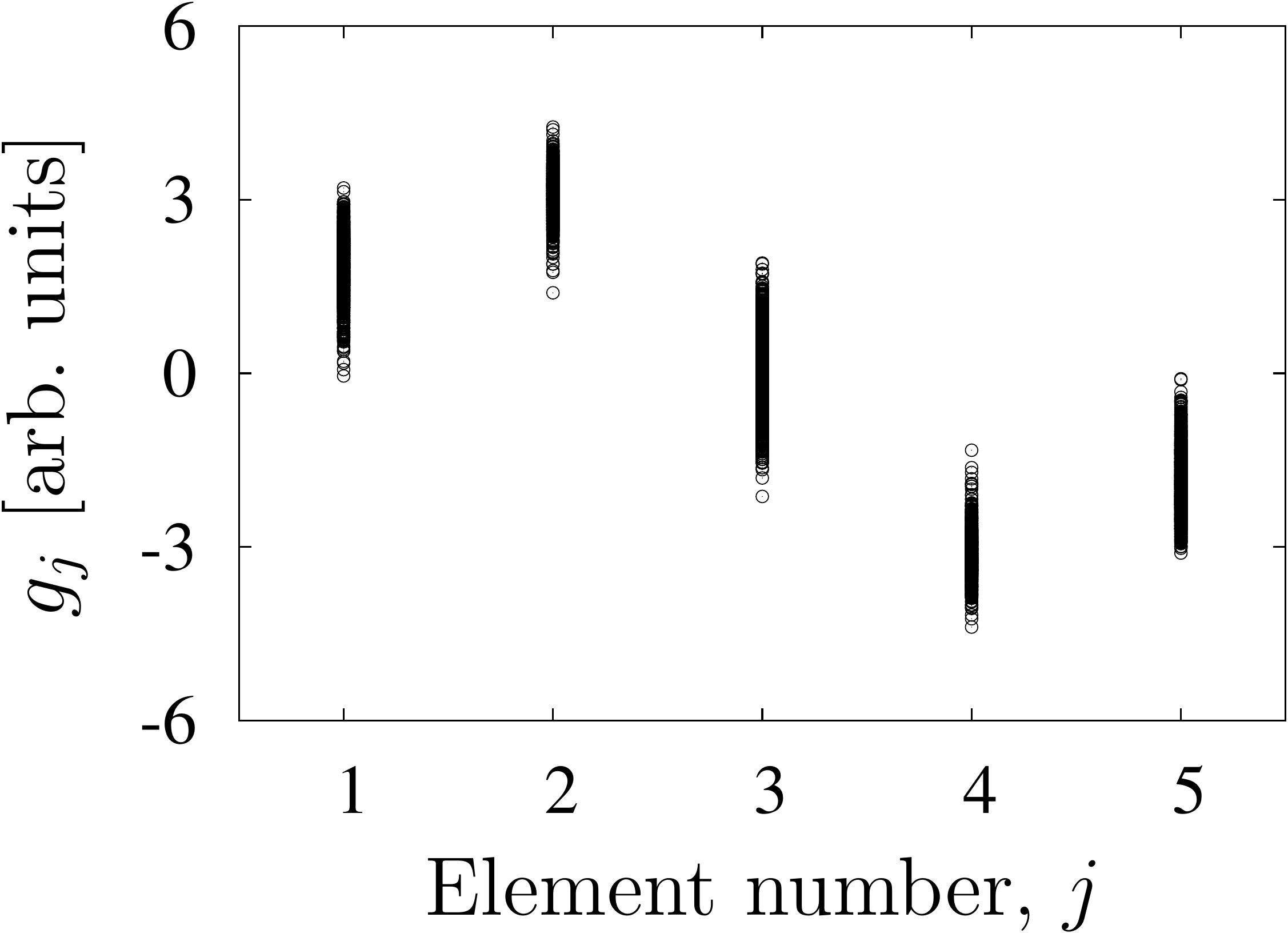}\quad\ 
 \includegraphics[width=0.5\figurewidth]{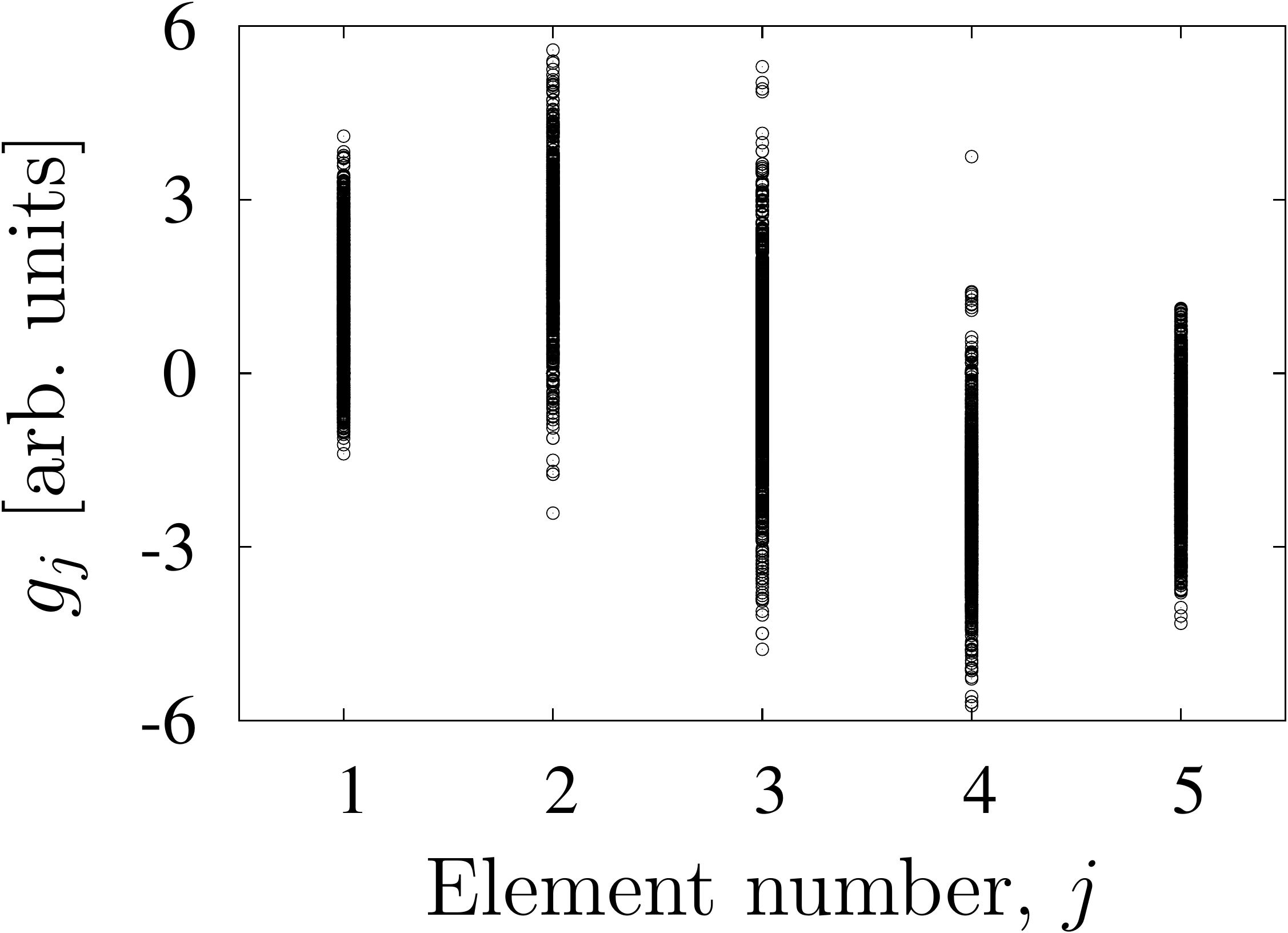}
\caption{Profiles of the `sinusoidal mode' (\ie, the coupling strength of each element, in arbitrary units) in a $5$-element array, for a representative sample of $1000$ populations, for a total of $4000$ populations, demonstrating random positioning errors of varying degree. The position of each element is varied from the ideal case by a random number drawn from a Gaussian distribution with standard deviation, from left to right, of $10^{-3}\lambda$, $10^{-2.5}\lambda$, $10^{-2}\lambda$, and $10^{-1.5}\lambda$. The overall coupling strength resulting from each of these simulations is shown in \fref{fig:dlOverallCouplings}. ($\zeta=-0.5$, $d=d_-$, other parameters as in the main text.)}
 \label{fig:dlIndividualCouplings}
\end{figure*}%
where $\alpha$ and $\beta$ are increments of first order in the relevant displacement [note that the (off-)diagonal terms are complex conjugates of each other; this is different to the case where absorption is nonzero]. When this matrix is substituted into the equation for the resonance condition, the terms involving $\re{e^{-i\mu}\alpha}$ and $\re{\beta}$ drop out entirely \emph{for a symmetric system}, such that it suffices to consider only the imaginary part of the increment. Let us reiterate that this happens only because the off-diagonal terms are complex conjugates of each other; were absorption to be nonzero, this would no longer be the case. \eref{eq:GeneratorEquation} now simplifies to
\begin{equation}
\frac{\partial k}{\partial\delta x_j}=-\frac{\im{\beta+e^{-i\mu}\alpha}}{L+2d\frac{\partial\chi}{\partial\nu}}\,,
\end{equation}
with $\nu=kd$,
\begin{align}
\alpha&=2ik\zeta\Bigl[e^{i\mu_1}(1+i\chi_1)\chi_2-e^{i\mu_2}\chi_1(1+i\chi_2)\Bigr]\nonumber\\
&=2ik\zeta^2\Biggl[\frac{(1+\zeta^2)U_{n_1-1}^2(a)U_{n_2-1}(a)}{(1-i\zeta)U_{n_1-1}(a)-e^{i\nu}U_{n_1-2}(a)}\nonumber\\
&\phantom{=2ik\zeta^2\Biggl[}\quad-\frac{(1+\zeta^2)U_{n_2-1}^2(a)U_{n_1-1}(a)}{(1-i\zeta)U_{n_2-1}(a)-e^{i\nu}U_{n_2-2}(a)}\Biggr]\,,
\end{align}
and
\begin{align}
\beta&=2k\zeta\Bigl[\chi_1\chi_2-(1+i\chi_1)(1-i\chi_2)e^{i(\mu_1-\mu_2)}\Bigr]\nonumber\\
&=2k\zeta \biggl\{\zeta^2U_{n_1-1}(a)U_{n_2-1}(a)\nonumber\\
&\phantom{=2k\zeta\biggl\{}\quad-\bigl[1+\zeta^2U_{n_1-1}^2(a)\bigr]\nonumber\\
&\phantom{=2k\zeta\biggl\{}\qquad\times\frac{(1-i\zeta)U_{n_2-1}(a)-e^{i\nu}U_{n_2-2}(a)}{(1-i\zeta)U_{n_1-1}(a)-e^{i\nu}U_{n_1-2}(a)}\biggr\}\,.
\end{align}
These two expressions simplify considerably to yield
\begin{multline}
\im{\beta+e^{-i\mu}\alpha}=2k\zeta\csc\Bigl(\frac{\pi}{N}\Bigr)\\
\times\biggl[\sqrt{\sin^2\Bigl(\frac{\pi}{N}\Bigr)+\zeta^2}-\zeta\biggr]\sin\biggl(2\pi\frac{j-\tfrac{1}{2}}{N}\biggr)\,,
\end{multline}
and therefore
\begin{multline}
g_j=-2\omega_\mathrm{c}x_0\frac{\zeta\csc\Bigl(\frac{\pi}{N}\Bigr)\biggl[\sqrt{\sin^2\Bigl(\frac{\pi}{N}\Bigr)+\zeta^2}-\zeta\biggr]}{L-2Nd\zeta\csc^2\bigl(\frac{\pi}{N}\bigr)\sqrt{\sin^2\bigl(\frac{\pi}{N}\bigr)+\zeta^2}}\\
\times\sin\biggl(2\pi\frac{j-\tfrac{1}{2}}{N}\biggr)\,.
\end{multline}

\begin{figure}[t]
 \includegraphics[width=\figurewidth]{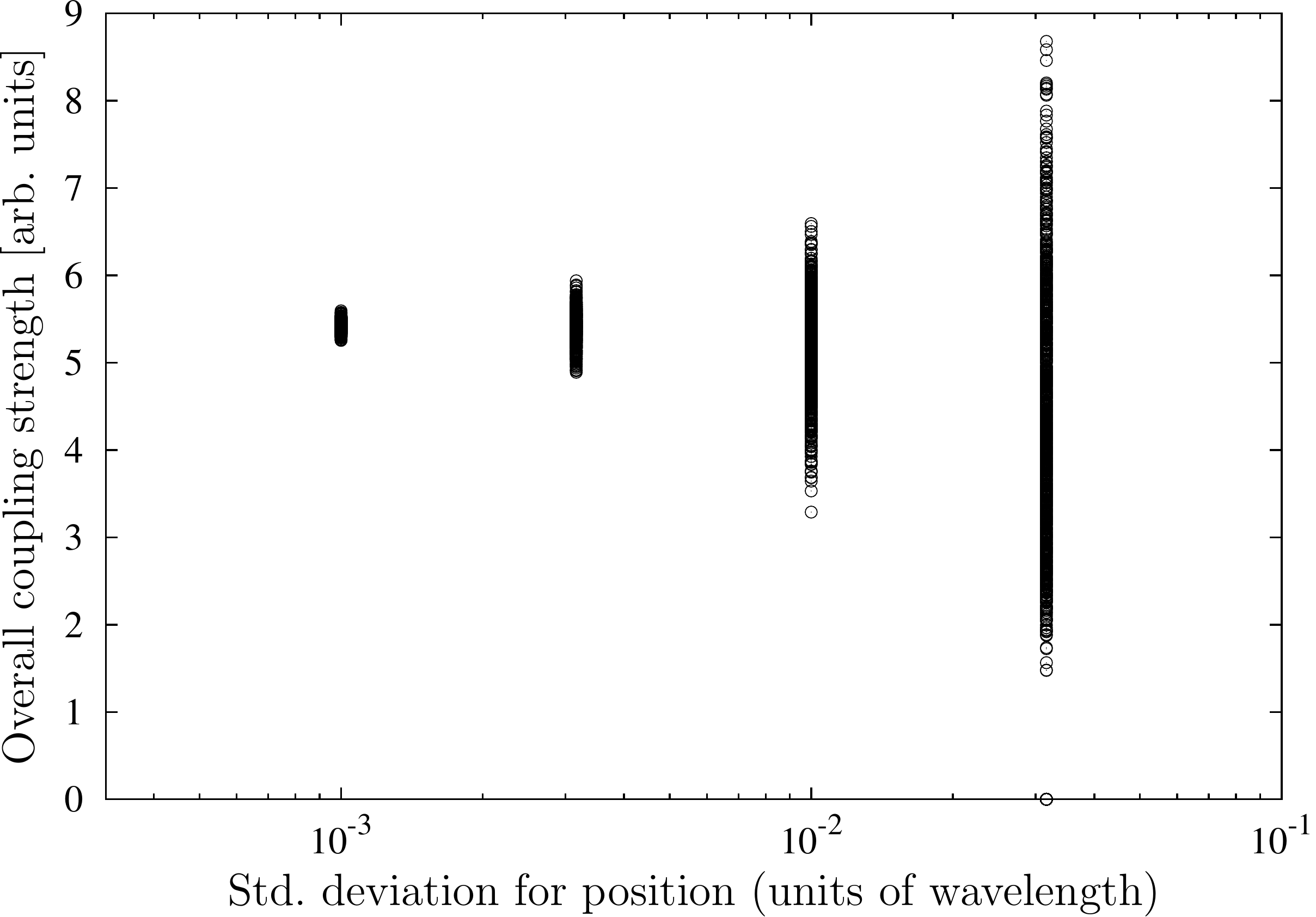}
\caption{Overall coupling strength for the populations shown in \fref{fig:dlIndividualCouplings}. Compared to the ideal case, the coupling strength for the four cases has an average (standard deviation) of $0.0$\% ($1.0$\%), $-0.3$\% ($3.0$\%), $-3.4$\% ($9.4$\%), and $-21.5$\% ($37.2$\%). (Parameters as in \fref{fig:dlIndividualCouplings}.)}
 \label{fig:dlOverallCouplings}
\end{figure}

\begin{figure*}[tb]
 \includegraphics[width=\figurewidth]{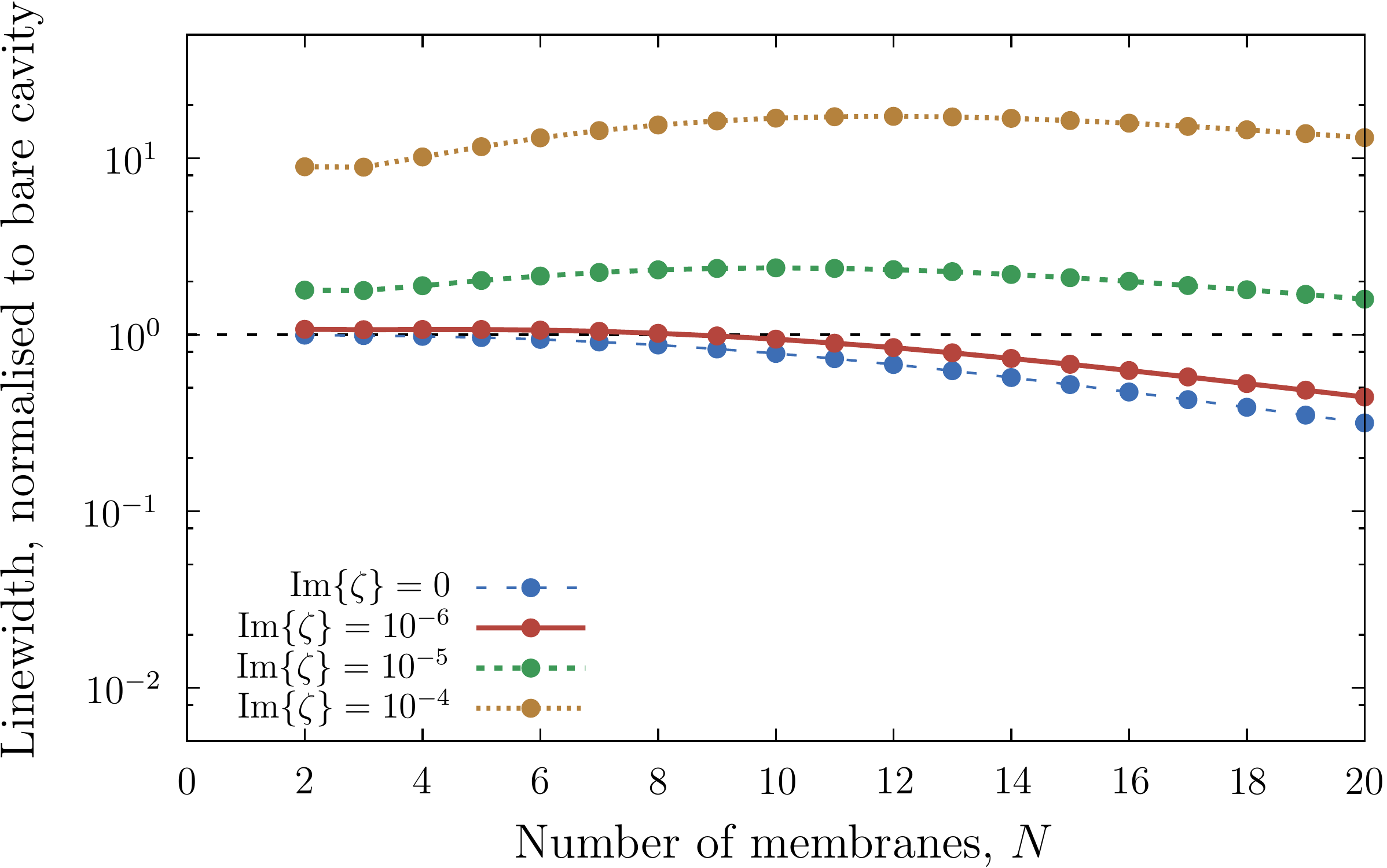}\quad
 \includegraphics[width=\figurewidth]{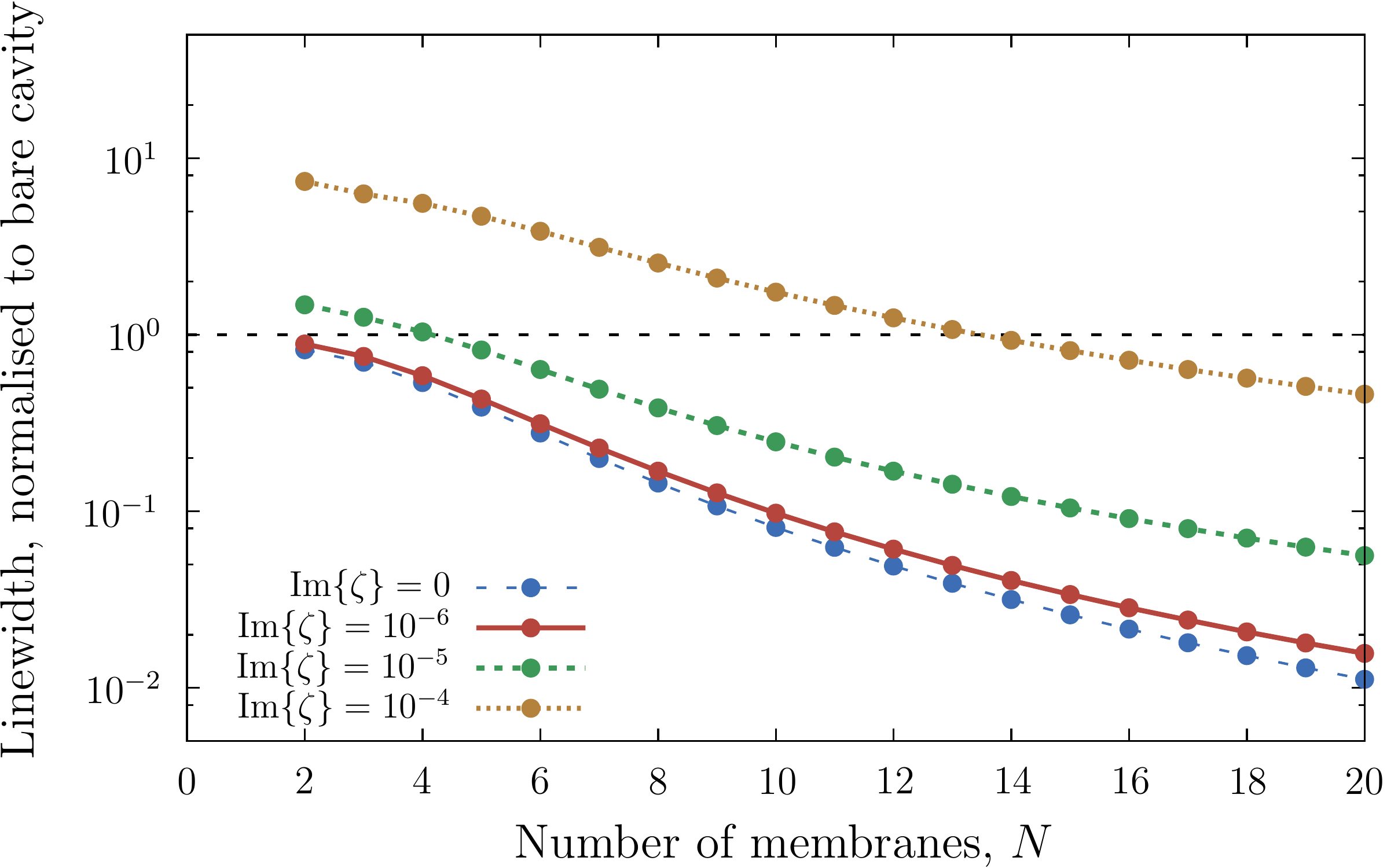}
\caption{Effect of increasing absorption for $\re{\zeta}=-12.9$ and $d=d_-$ (left), $d=d_-+20\lambda$ (right). Larger inter-element separations make the system more tolerant to higher levels of absorption. The linewidth of the bare cavity is represented by the horizontal dashed black lines. These figures should be compared to Fig.\ 4 of the main text. (Bare cavity finesse $\approx\,3\times10^4$, other parameters as in the main text.)}
 \label{fig:Absorption}
\end{figure*}

As discussed in the main text, the coupling of the collective motion of the membranes to the cavity field is governed by the constant $\sqrt{\sum_{j=1}^Ng_j^2}$, such that for $N>2$
\begin{equation}
g_\mathrm{sin}=-\sqrt{\frac{N}{2}}\frac{g\,\zeta\csc\bigl(\frac{\pi}{N}\bigr)\Bigl[\sqrt{\sin^2\bigl(\frac{\pi}{N}\bigr)+\zeta^2}-\zeta\Bigr]}{1-2N\tfrac{d}{L}\zeta\csc^2\bigl(\frac{\pi}{N}\bigr)\sqrt{\sin^2\bigl(\frac{\pi}{N}\bigr)+\zeta^2}}
\end{equation}
because of the relation
\begin{equation}
\sqrt{\sum_{j=1}^N\sin^2\biggl(2\pi\frac{j-\tfrac{1}{2}}{N}\biggr)}=\begin{cases}
\sqrt{2} & \text{for }N=2\\
\sqrt{\frac{N}{2}} & \text{for }N>2
\end{cases}\,;
\end{equation}
this is equal to $g_\mathrm{sin}$ as defined in the main text in the appropriate limits.

\section{Tolerance of mechanism to imperfections}
Our model makes three key assumptions about the physical make-up of the membrane ensemble, which we will now discuss in brief. The membranes are assumed to be `thin'; since silicon nitride membranes have thicknesses of ca.\ $50$\,nm~\cite{Thompson2008}, this is not expected to be a limitation. The reason behind this limitation is technical rather than practical:\ we assume that the reflectivity of each element in the array is independent of frequency. For atoms, this means that the detuning of the cavity field from resonance is much larger than the atomic linewidth. In the case of membranes the requirement is similar, in that we need the polarisability to be effectively constant with respect to frequency. Let us model each membrane as a plate of (real) refractive index $n$ and thickness $l$ (see \fref{fig:Plate}). Going from left to right, the two interfaces have amplitude reflectivities $\mathfrak{r}_1=\mathfrak{r}=(1-n)/(1+n)$ and $\mathfrak{r}_2=-\mathfrak{r}$, as well as transmissivities $\mathfrak{t}_1=2/(1+n)$ and $\mathfrak{t}_2=2n/(1+n)$, which are obtained from the respective Fresnel coefficients at normal incidence. The amplitude reflection and transmission coefficients for light of wavenumber $k$ incident on the plate as a whole can then be written
\begin{equation}
r=\frac{\mathfrak{r}\bigl(1-e^{-2inkl}\bigr)}{1-\mathfrak{r}^2e^{-2inkl}}\,,\text{\ and\ }t=\frac{\mathfrak{t}_1\mathfrak{t}_2e^{-inkl}}{1-\mathfrak{r}^2e^{-2inkl}}\,,
\end{equation}
respectively. We then define the effective membrane polarizability $\zeta\equiv-ir/t$~\cite{Deutsch1995}, whereupon
\begin{equation}
\zeta=\frac{2\mathfrak{r}}{\mathfrak{t}_1\mathfrak{t}_2}\sin(nkl)=\frac{1-n^2}{2n}\sin(nkl)\,.
\end{equation}
Let us calculate the relative change in $\zeta$ over an interval of size $\kappa_\mathrm{c}$:
\begin{equation}
\frac{\partial\zeta/\partial k}{\zeta}\Delta k=\frac{nkl}{\tan(nkl)}\frac{1}{Q_\mathrm{c}}\,,
\end{equation}
where $Q_\mathrm{c}=\omega_0/\kappa_\mathrm{c}$ is the quality factor of the cavity, which is generally of the order of $10^7$--$10^9$. As shown below, our results are robust to well beyond perturbations in $\zeta$ of order $1/Q_\mathrm{c}$. For membranes that are thin on the scale of a wavelength, this requirement is thus satisfied trivially; for thicker membranes one must take care to avoid the resonances at $\sin(nkl)=0$. Another reason for requiring the membranes to be thin is that the free-spectral range of each `membrane-etalon' must be significantly larger than that of the main cavity, which requires $l\lll L/n$. A final, and stronger, requirement is that the entire membrane stack should fit within the Rayleigh range of the cavity.
\par
Positioning errors are also a concern. We performed numerical simulations with several thousand arrays ($N=5$, $\zeta=-0.5$, $L\approx6.3\times10^4\lambda$, $d=d_-$) whose elements were each shifted from the optimal position by a random shift drawn from a Gaussian distribution with a standard deviation of $10$\,nm. The resulting coupling strengths were within $\pm10$\% of the analytically-calculated value. A representative sample of the analysis performed on $4000$ arrays is shown in \fref{fig:dlIndividualCouplings} and \fref{fig:dlOverallCouplings}.
\par
The membranes were also assumed to be identical and non-absorbing; the latter is an excellent approximation to a single membrane with an imaginary part of the refractive index being $\lesssim10^{-6}$--$10^{-5}$~\cite{Wilson2009,*Karuza2011}. Fluctuations in the polarizability of the individual elements by up to $\pm10$\% give coupling strengths within $\pm6$\% of the analytic value. Our numerical investigations show that the major effect of absorption is not on the optomechanical coupling strength, but on the linewidth of the cavity. High levels of absorption would therefore not alter the coupling strength significantly, but would limit the achievable cavity finesse. A systematic study of the effect of absorption on the cavity linewidth is shown in \fref{fig:Absorption}. The cavity linewidth, both in the presence of absorption and in its absence, is calculated numerically by scanning over, and fitting a Lorentzian to, the cavity resonance. We are not aware of a concise analytic formulation that would allow us to estimate the linewidth more directly for our situation.\\
To stay within the frame of the 1D model considered here, a small misalignment in the individual elements could also be simply modeled similarly to absorption (\ie, through a nonzero $\im{\zeta}$), since both effects represent a loss channel for the cavity field. Other detrimental effects of absorption, such as heating, are mitigated by the large coupling strengths obtained, which allow much smaller photon numbers to be used ($g_\mathrm{sin}^2\varpropto N^3$ increases faster than the absorbed power as $N$ increases). We note also that at large input powers it might be possible to exploit photothermal forces to further enhance the collective optomechanical interaction~\cite{Pinard2008,*Restrepo2011,Metzger2004,*Usami2012,*Xuereb2012b}.

\begin{figure}[t]
 \includegraphics[width=\figurewidth]{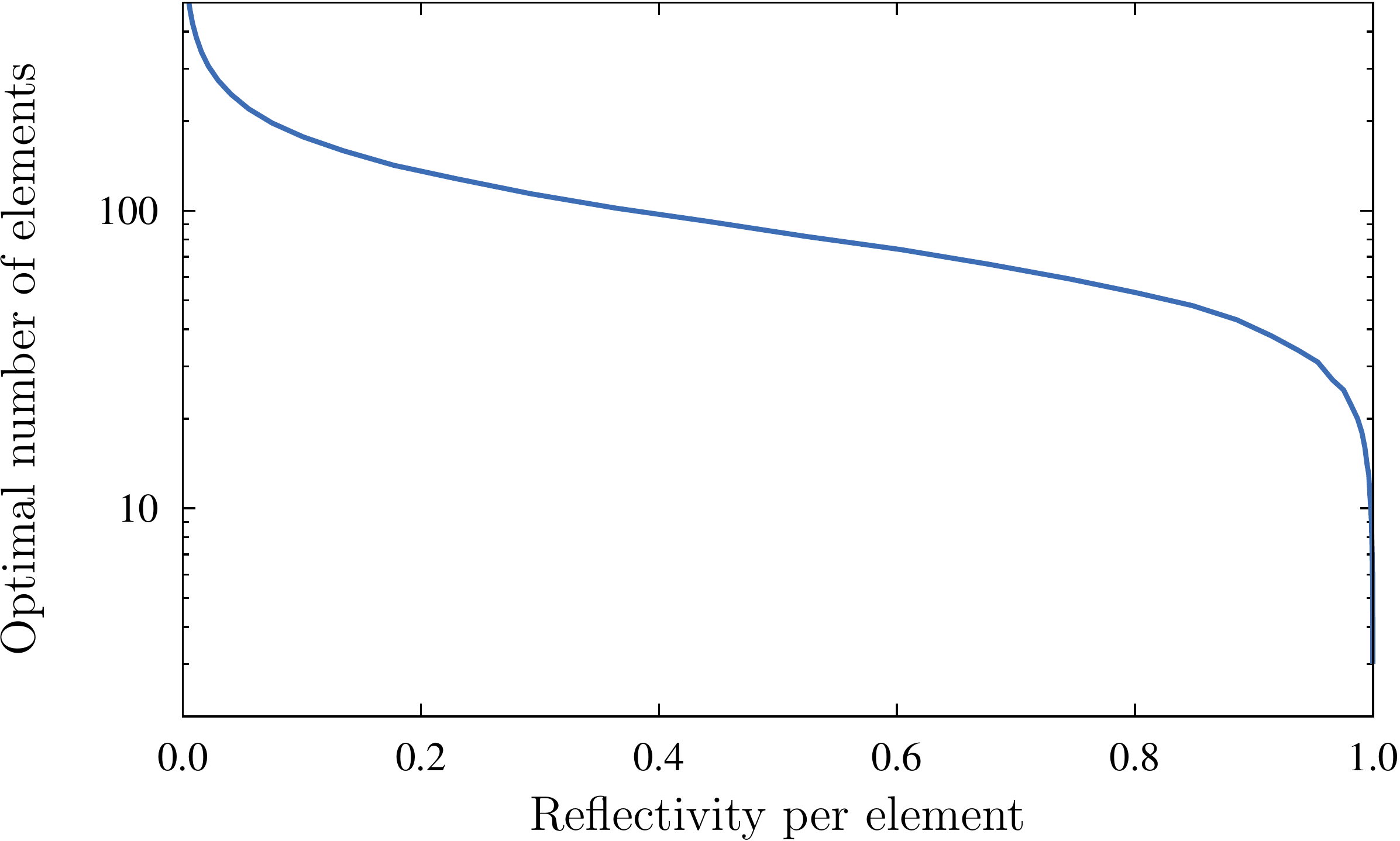}
\caption{Optimal number of elements, $N_\mathrm{opt}$, as a function of the single-element reflectivity; for low values of the reflectivity, $g_\mathrm{sin}$ rises less steeply with $N$ and is therefore close to its optimum value even for much larger $N$. ($\zeta=-0.5$, $d=d_-$, other parameters as in the main text.)}
 \label{fig:Nopt}
\end{figure}
\section{Optimal number of elements}
In the main text we optimized $g_\mathrm{sin}$ over $N$ to obtain the maximal coupling strength $g_\mathrm{sin}^\mathrm{opt}$. The value of $N$ for which $g_\mathrm{sin}=g_\mathrm{sin}^\mathrm{opt}$ is shown as a function of the per-element reflectivity in \fref{fig:Nopt}. The lower the per-element reflectivity, the weaker the dependence of $g_\mathrm{sin}$ is on $N$; indeed, in such cases $N_\mathrm{opt}$ is very large, but $g_\mathrm{sin}$ is close to its optimal value for much smaller values of $N$. These large values of $N_\mathrm{opt}$ can be looked at, from a different point of view, as witnesses of the fact that the mechanism we describe does not saturate quickly with increasing $N$.

\begin{figure}[t]
 \includegraphics[width=\figurewidth]{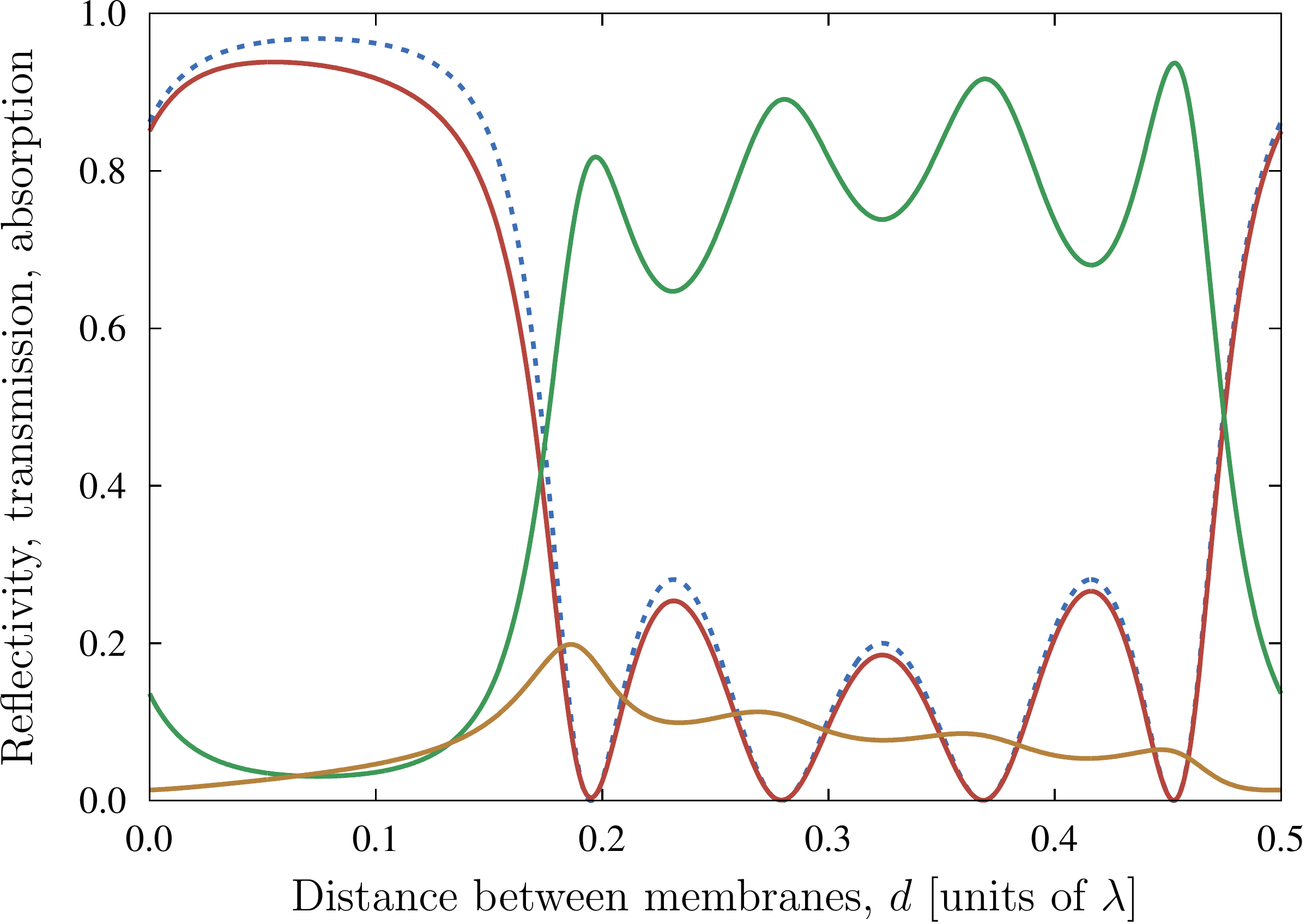}
\caption{Reflectivity (red), transmission (green), and absorption (orange) for $N=5$ elements with an individual reflectivity of ca.\ $50$\% and a very large absorption. The dashed blue curve is identical to the blue curve in Fig.\ 1(c) of the main text. Note that the absorption is highest close to $d_+$ and lowest close to $d_-$.}
 \label{fig:AbsorptionScan}
\end{figure}
\section{Absorption at $d_\pm$}
For a nonzero per-element absorption, the largest amount of absorption appears close to the points where the ensemble is transparent. This is shown in \fref{fig:AbsorptionScan}, where we plot the reflectivity, transmission, and absorption of an ensemble of $5$ elements as the spacing between the elements is scanned. This figure is meant to complement Fig.\ 1(c) in the main text. A general feature is that the absorption is largest at $d_+$ and rather smaller at $d_-$; we therefore choose the latter as our working point in the text.

\section{Independence of mechanical decay rate on $N$}
Let us describe the motion of the $j$\textsuperscript{th} mechanical element ($1\leq j\leq N$) through the annihilation operator $\hat{b}_j$, which obeys the Heisenberg--Langevin equation of motion
\begin{equation}
\tfrac{\rmd}{\rmd t}\hat{b}_j=-(i\omega_\mathrm{m}+\Gamma)\hat{b}_j+\hat{F}_j+\sqrt{2\Gamma}\hat{\xi}_j\,,
\end{equation}
where $\hat{\xi}_j$ is the relevent Langevin noise term. For simplicity, assume that all the oscillators have identical oscillation frequency $\omega_\mathrm{m}$, decay rate $\Gamma$, and temperature $T$, such that in thermal equilibrium they all have the same average occupation. $\hat{F}_j$ is a force term, perhaps due to the action of the cavity, whose form is not relevant here. To describe the collective motion, we use the vector $(g_j)$, normalized such that $\sum_{j=1}^N{g_j^2}=1$, and define:\ $\hat{b}=\sum_{j=1}^N{g_j\hat{b}_j}$ and $\hat{\xi}=\sum_{j=1}^N{g_j\hat{\xi}_j}$. Thus:
\begin{equation}
\tfrac{\rmd}{\rmd t}\hat{b}=-(i\omega_\mathrm{m}+\Gamma)\hat{b}+\sum_{j=1}^N{g_j\hat{F}_j}+\sqrt{2\Gamma}\hat{\xi}\,.
\end{equation}
Under the assumption that the noise terms $\hat{\xi}_j$ are of a similar nature to one another and are independent (i.e., any cross-correlator between $\hat{\xi}_i$ and $\hat{\xi}_j$ is zero for $i\neq j$), then $\hat{\xi}$ obeys the same correlation functions as each \emph{individual} noise term, because of the normalisation of $(g_j)$, whereupon $\hat{b}$ behaves as a single collective oscillator with decay rate $\Gamma$.\\
Finally, let us remark that our description in terms of this collective mode is one where we merely rotate to a different basis for this $N$-dimensional space, and therefore the correct normalisation, necessary for the rotation to be a unitary operation, is $\sum_{j=1}^N{g_j^2}=1$.

\begin{figure*}[t]
 \includegraphics[width=0.65\figurewidth]{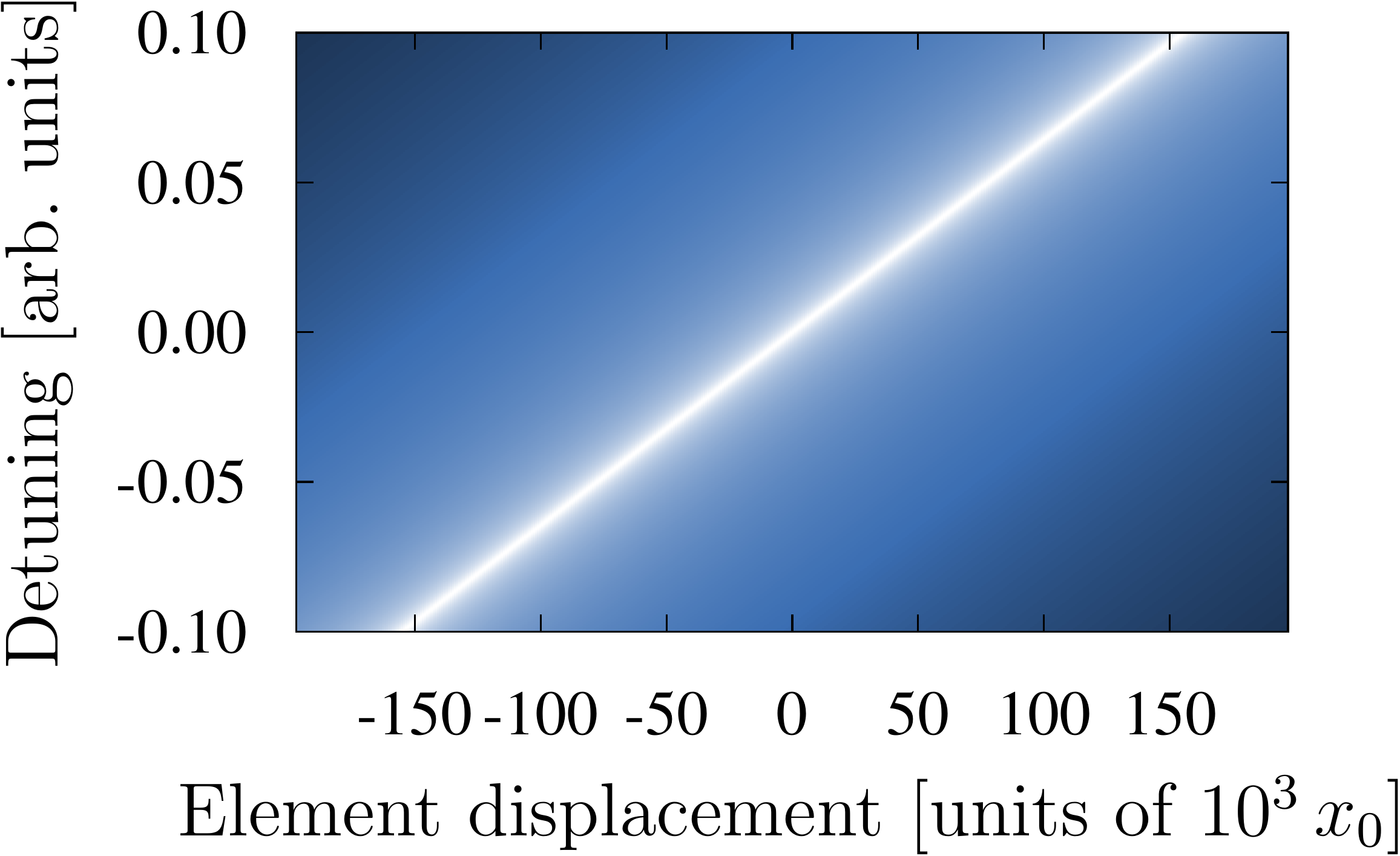}\quad\ 
 \includegraphics[width=0.65\figurewidth]{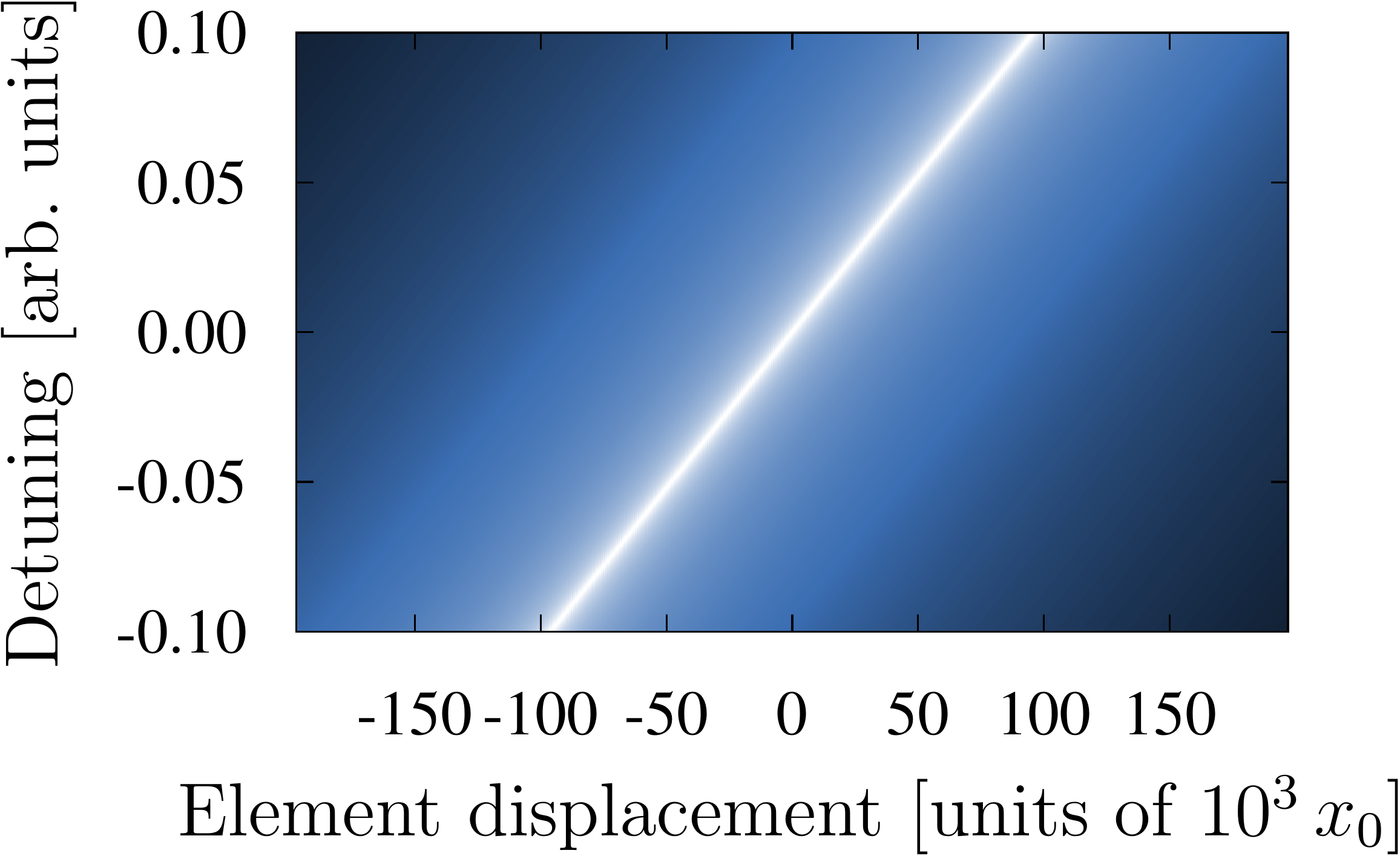}\quad\ 
 \includegraphics[width=0.65\figurewidth]{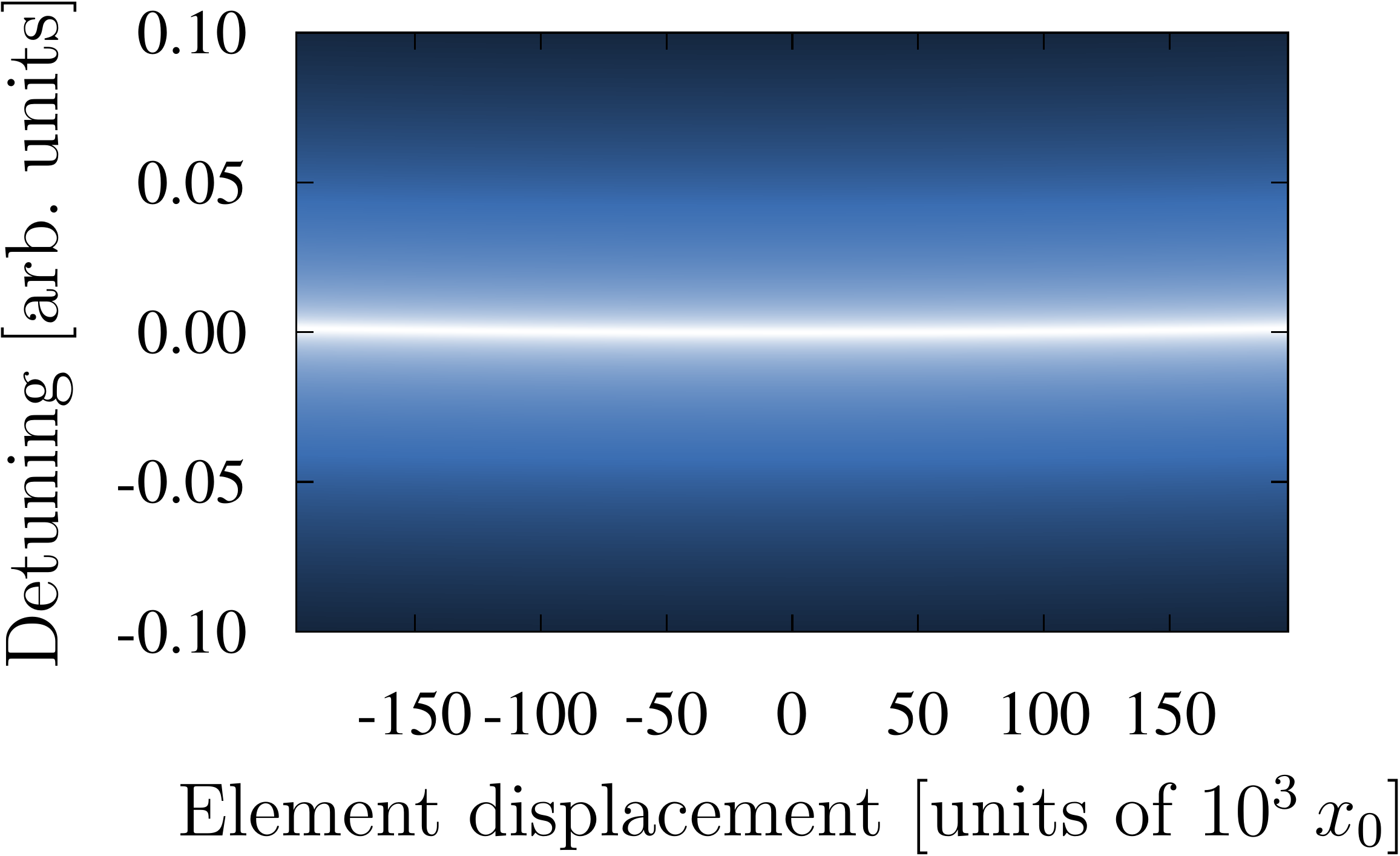}\\[5mm]
 \includegraphics[width=0.65\figurewidth]{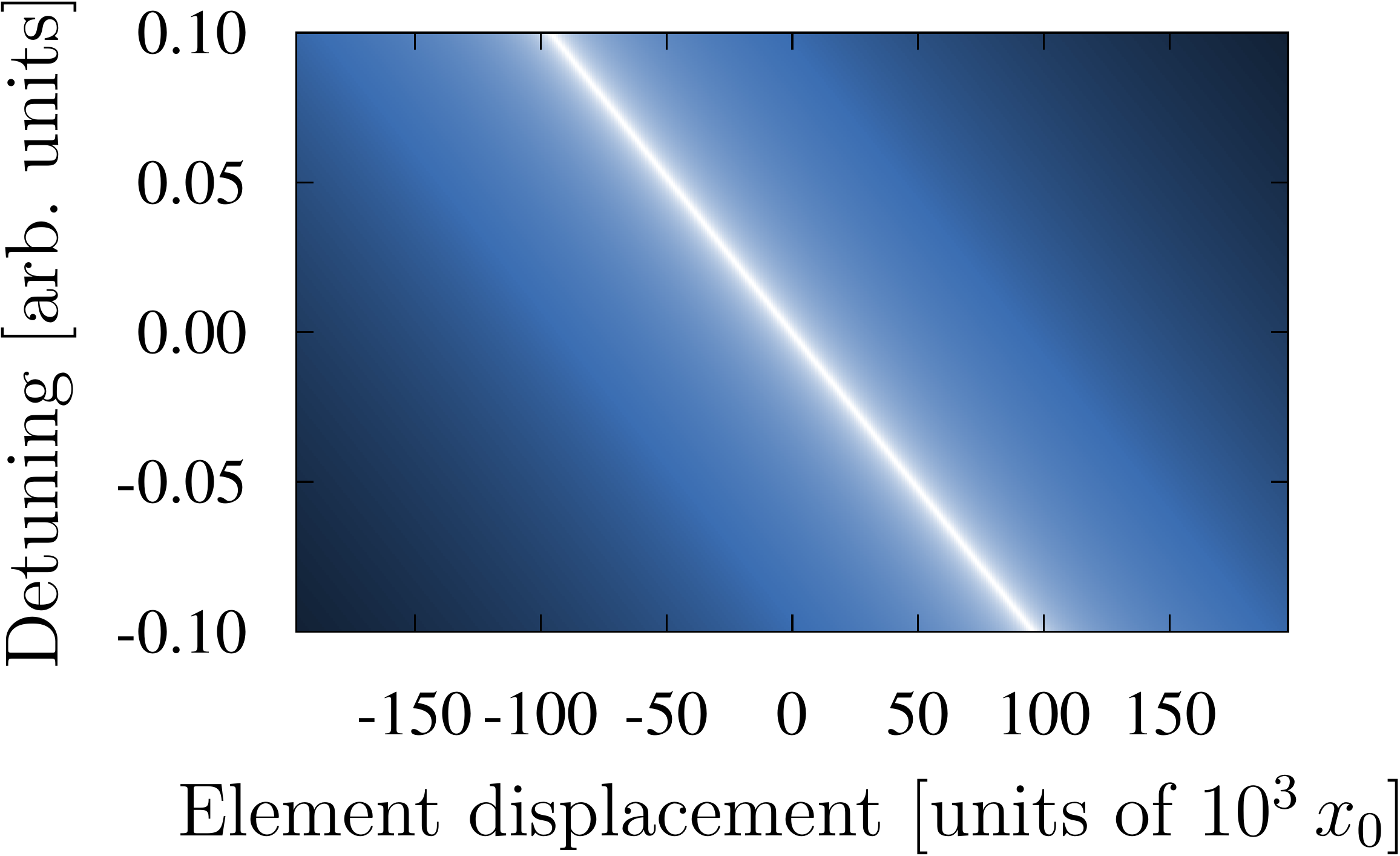}\quad\ 
 \includegraphics[width=0.65\figurewidth]{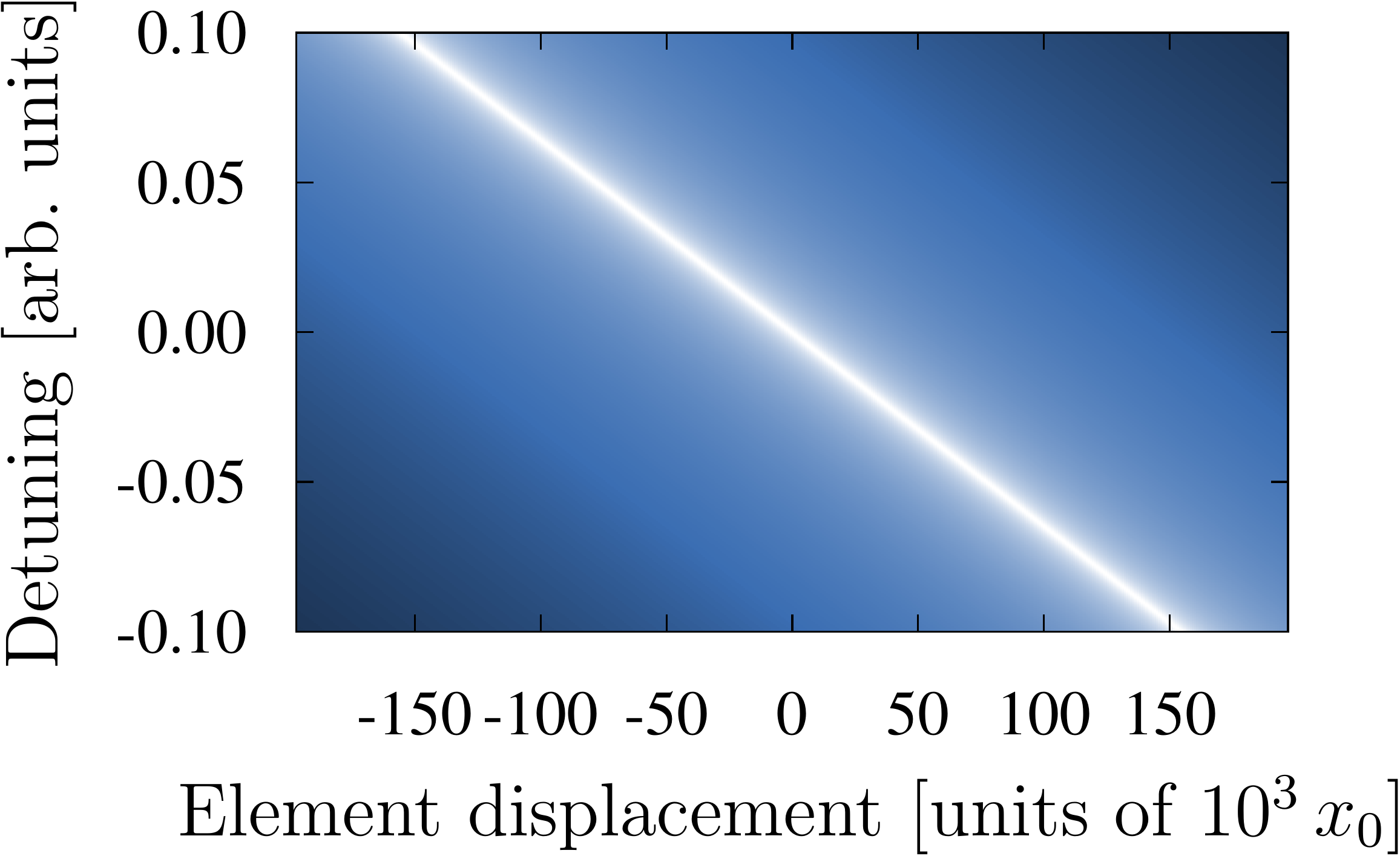}
\caption{Transmission plots that show the variation of the cavity resonance frequency as each element in the ensemble is displaced. For this set of data we have $N=5$, $\zeta=-0.5$, $L\approx0.25$\,cm, $x_0=2.7$\,fm. In the top row we have $j=1$, $j=2$, and $j=3$ (from left to right), and in the second row $j=4$ and $j=5$. For the center element ($j=3$), the slope of the curve around the rest position is zero (\ie, $g_3=0$).}
 \label{fig:ElementsLow}
\end{figure*}%
\begin{figure*}[t]
 \includegraphics[width=0.65\figurewidth]{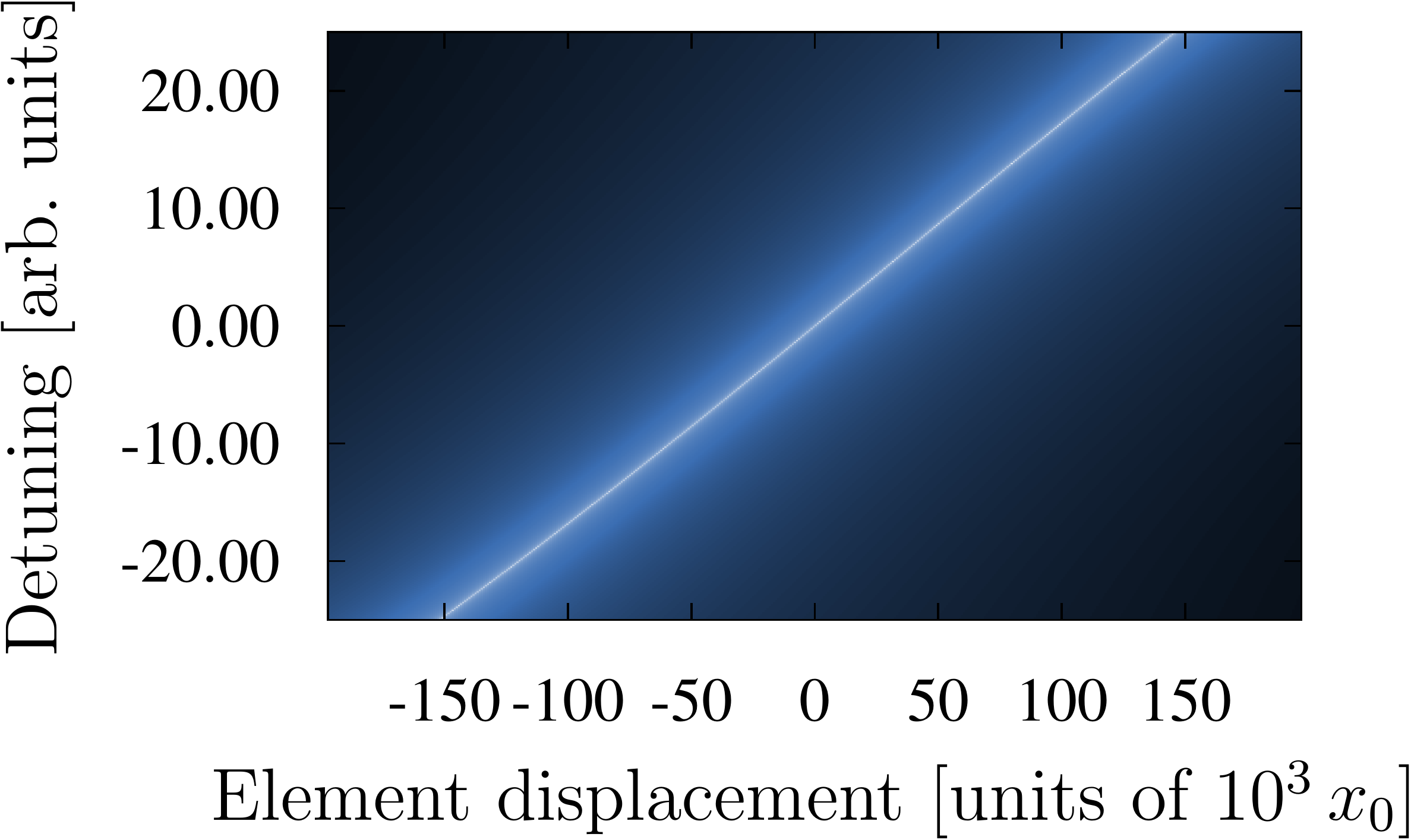}\quad\ 
 \includegraphics[width=0.65\figurewidth]{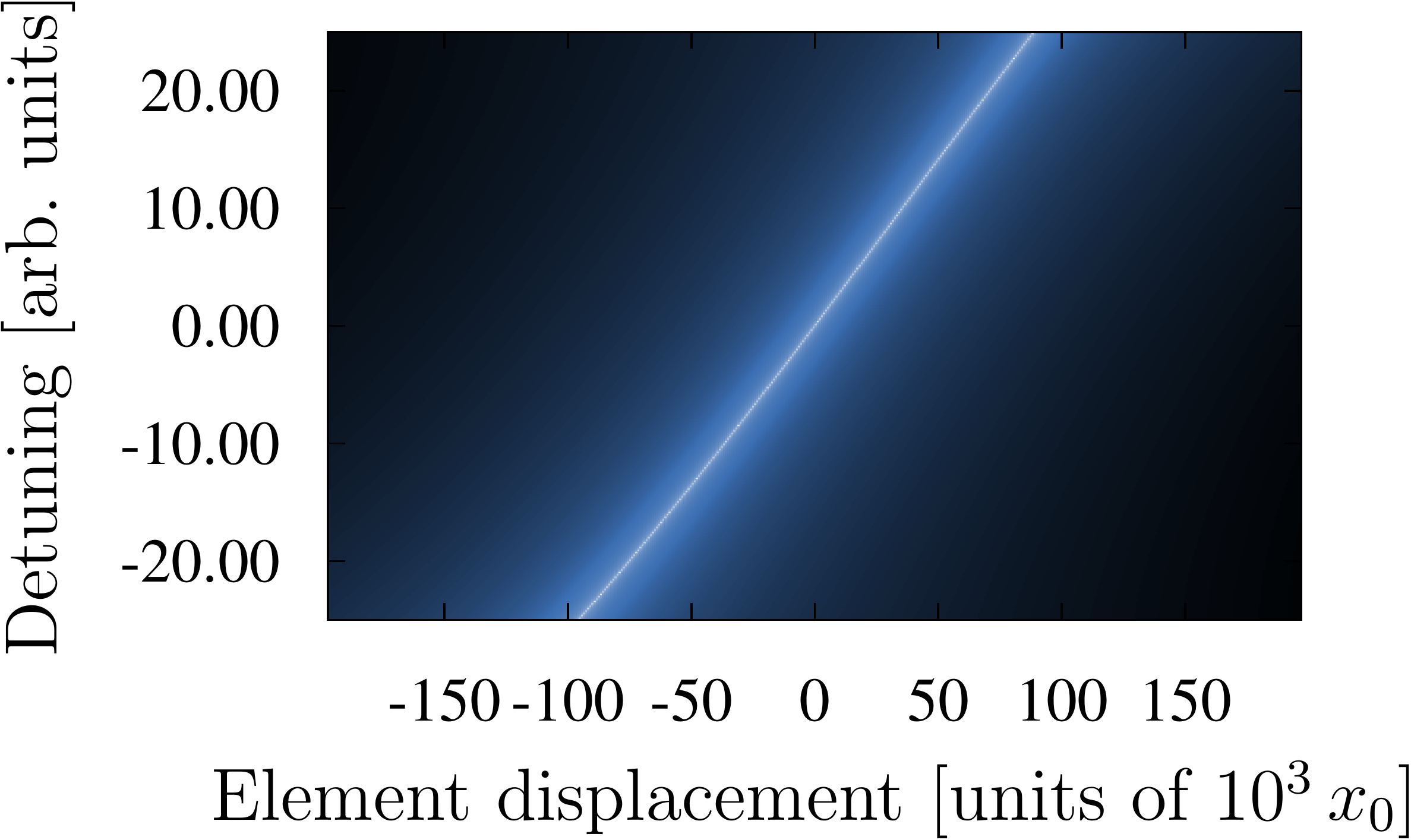}\quad\ 
 \includegraphics[width=0.65\figurewidth]{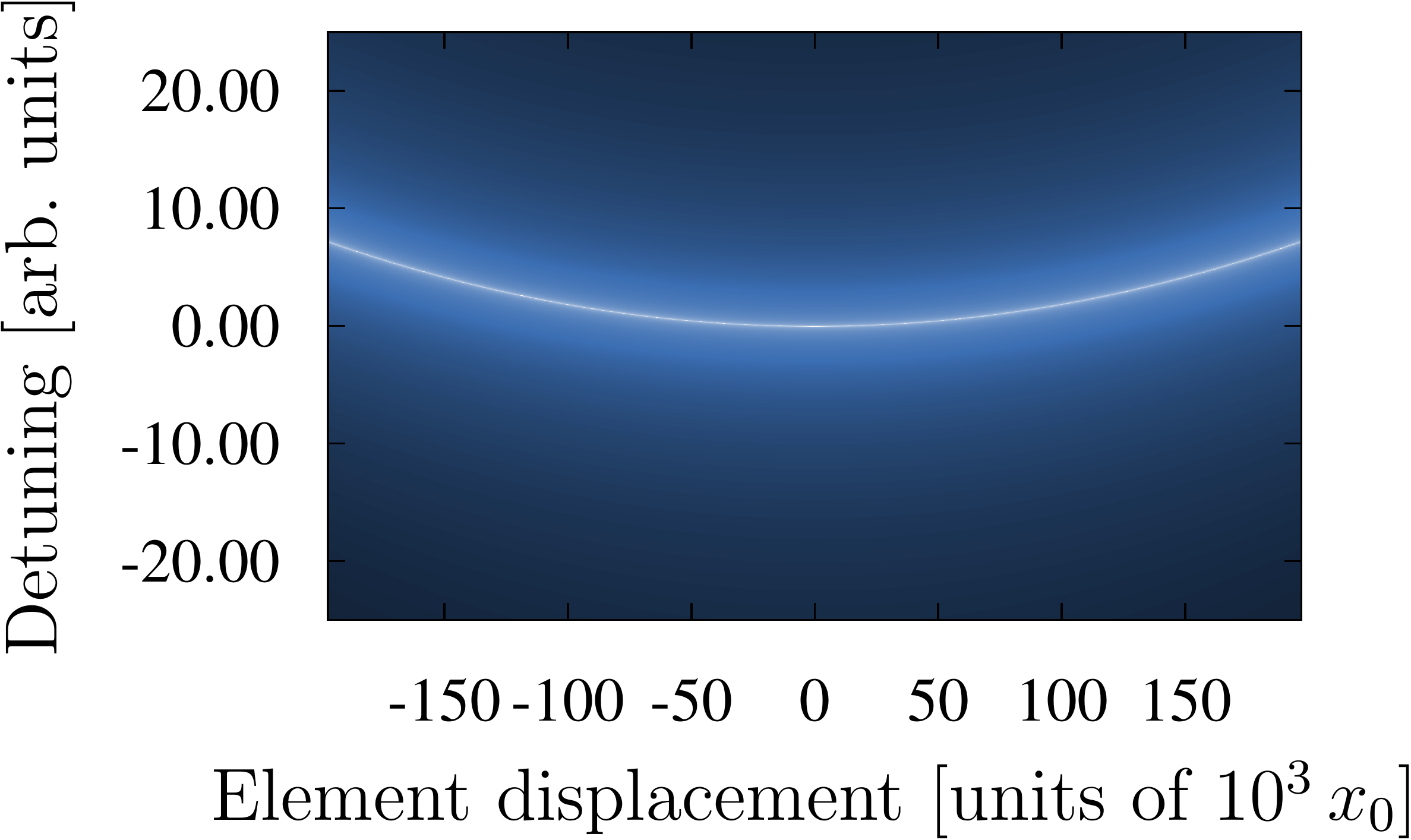}\\[5mm]
 \includegraphics[width=0.65\figurewidth]{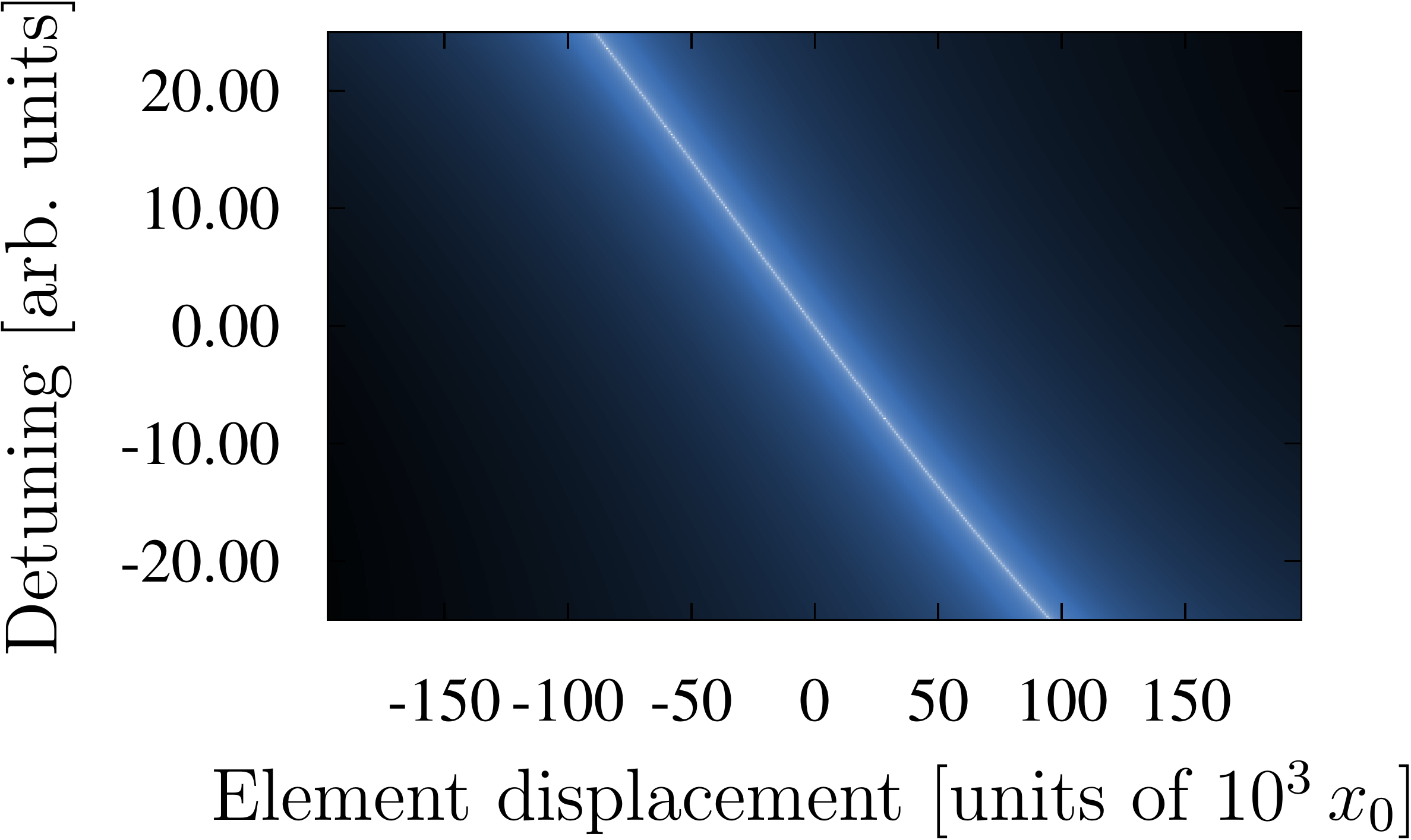}\quad\ 
 \includegraphics[width=0.65\figurewidth]{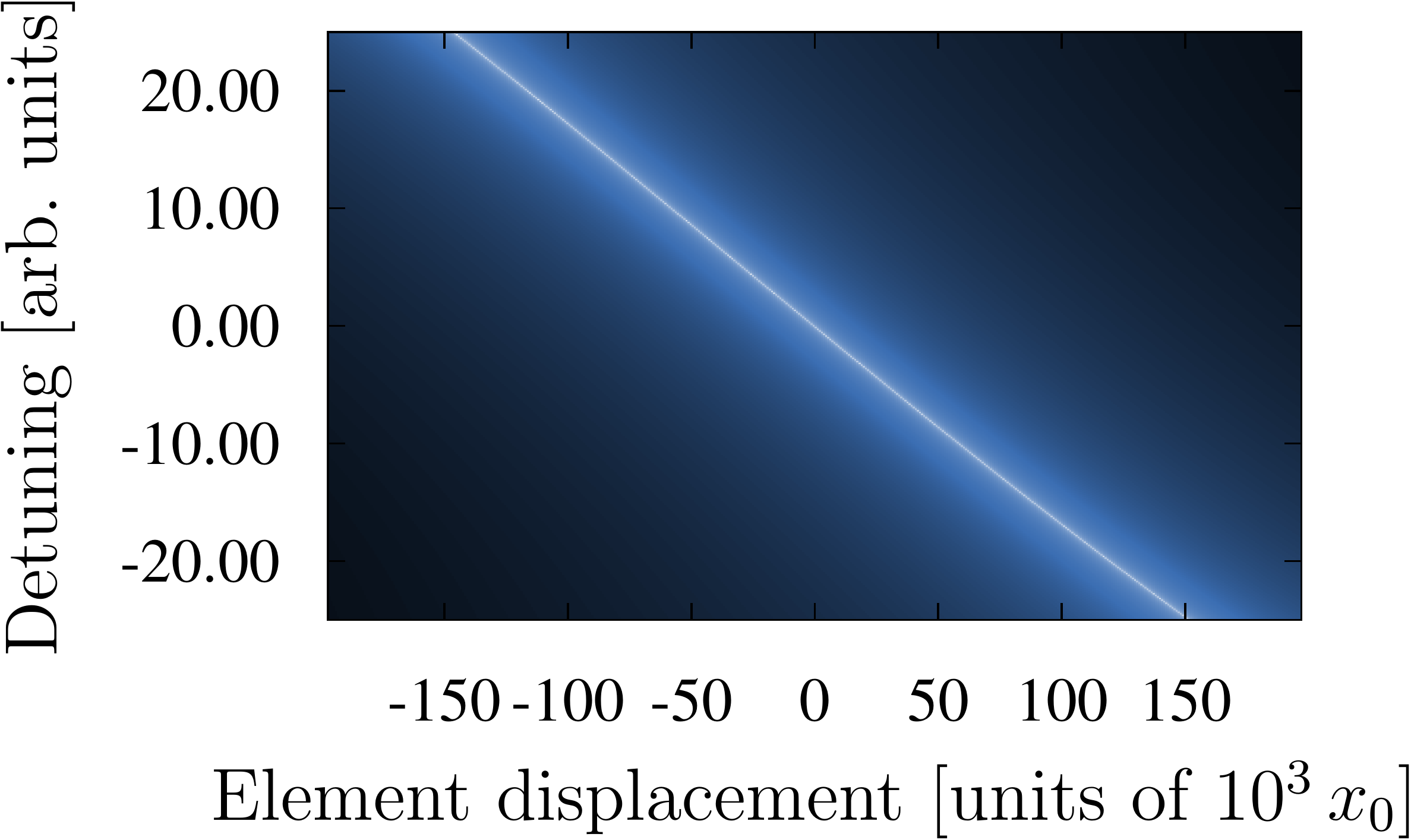}
\caption{Similar to \fref{fig:ElementsLow}, but with $\zeta=-12.9$. The scale on the vertical axis matches that in \fref{fig:ElementsLow}.}
 \label{fig:ElementsHigh}
\end{figure*}%
\section{Approximation of linear coupling to the motion}
Most of the results presented in the main text, especially those for arrays of identical lossless elements, were obtained analytically. These results were then confirmed by numerical methods, as detailed elsewhere in this manuscript, and extended to cases of lossy elements, non-uniform arrays, etc. In each case, the coupling constants $g_j$ were calculated by displacing the $j$\textsuperscript{th} element by an amount $\delta x_j$, thus shifting the resonance frequency of the cavity from $\omega$ to $\omega+\delta\omega$. Assuming a linear variation of $\omega$ with the positions of the elements, this gave us a numerical estimate for $g_j=(\delta\omega/\delta x_j)x_0$; the multiplication by $x_0$ serves to express $g_j$ as a frequency. The approximation of linear variation of $\omega$ with $x_j$ may break down in some situations~\cite{Thompson2008,Sankey2010}, but its validity in any one situation is fairly easy to check. Consider, for example, the data shown in \fref{fig:ElementsLow}. This is a set of plots showing the intensity of the light transmitted through the cavity as a function of the displacement of each element for a system with $N=5$. One notices that, even for displacements of the order of $10^5$ times the size of the zero-point fluctuations, the resonance frequency varies linearly with the position of all the membranes but the center one; indeed for $j=3$ and $N=5$ we obtain $g_j=0$.\\
This effect can be seen more clearly if we use very highly-reflective membranes; cf.\ \fref{fig:ElementsHigh}. We note that for $j=3$, the frequency depends quadratically on the coordinate $x_j$. Moreover, for $\zeta=-12.9$ and at a displacement of $\sim10^5\,x_0$, the frequency shift for $j=3$ is about an order of magnitude smaller than that for $j=1$. To estimate the effect of this quadratic dependence, let us define the quadratic optomechanical coupling strength $G_3=(\delta\omega/\delta x_3^2)x_0^2$, which has the units of frequency. We have already noted that
\begin{equation}
g_1\bigl(10^5\bigr)\sim10\,G_3\bigl(10^5\bigr)^2\,,\text{\ \ie,\ }\,G_3\sim10^{-6}g_1\,.
\end{equation}
In the situation considered in the main text, the linear coupling thus dominates largely over the quadratic coupling.\\Indeed, for the temperature and frequency given there ($T=1$\,K, $\omega_\mathrm{m}=2\pi\times211$\,kHz), the average occupation number is about $10^5$, which translates to a root-mean-squared displacement of ca.\ $300\,x_0$, and therefore a ratio for the quadratic to the linear frequency shift of about $3\times10^{-4}$.

\end{document}